\documentclass[11pt]{article}
\usepackage[usenames,dvipsnames,svgnames,table]{xcolor} 

\usepackage{enumitem} 
\usepackage{rotfloat} 
\usepackage{graphicx}
\usepackage{mathrsfs,textcomp,manyfoot,listings,array}
\usepackage{colortbl}
\usepackage{epsfig}
\usepackage{amssymb, amsmath}
\usepackage{verbatim}
\usepackage{natbib}
\usepackage{authblk}
\IfFileExists{kotex.sty}{\usepackage{kotex}}{} 
\usepackage{enumitem}
\usepackage{multirow}
\usepackage{booktabs} 
\usepackage{adjustbox}
\usepackage[colorlinks=true,linkcolor=blue,citecolor=blue,urlcolor=blue]{hyperref}%
\usepackage[colorinlistoftodos, textsize=scriptsize]{todonotes} 
\usepackage{subcaption}
\usepackage{algorithm}
\usepackage{algpseudocode}
\usepackage[titletoc,title]{appendix}
\usepackage{verbatim}
\usepackage{longtable}
\usepackage{hhline}

\newtheorem{theorem}{Theorem}
\newtheorem{lemma}[theorem]{Lemma}

\newtheorem{corollary}{Corollary}[theorem]

\newenvironment{proof}[1][Proof]{\begin{trivlist}
\item[\hskip \labelsep {\bfseries #1}]}{\end{trivlist}}
\newtheorem{definition}{Definition}
\newenvironment{remark}[1][Remark]{\begin{trivlist}
\item[\hskip \labelsep {\bfseries #1}]}{\end{trivlist}}

\newcommand{\bea}{\begin{eqnarray*}}
\newcommand{\eea}{\end{eqnarray*}}
\newcommand{\bean}{\begin{eqnarray}}
\newcommand{\eean}{\end{eqnarray}}

\newcommand{\bx}{\textbf{x}}

\newcommand{\pkg}[1]{\texttt{#1}}

\begin{document}

\title{Multivariate L{\'e}vy Adaptive B-Spline Regression}
\author[1]{Sewon Park}
\author[2]{Jaeyong Lee}
\author[2]{Jeunghun Oh}
\affil[1]{Department of Statistics, Sookmyung Women's University,
100 Cheongpa-ro 47-gil, Yongsan-gu, Seoul 04310, Republic of Korea}
\affil[2]{Department of Statistics, Seoul National University,
1 Gwanak-ro, Gwanak-gu, Seoul 08826, Republic of Korea}
\date{}

\maketitle

\begin{abstract}
We propose Multivariate L\'evy Adaptive B-Spline Regression (MLABS), a Bayesian nonparametric regression model designed to capture nonlinear effects, interactions, and locally varying features of multivariate functions.
MLABS represents the regression function as a random sum of low-order
tensor-product B-spline basis functions under a finite L\'evy random
measure prior. The resulting representation adapts to the complexity
and local structure of the regression function while favoring a sparse
basis representation. For posterior inference, we develop a reversible-jump Markov chain Monte Carlo (MCMC) algorithm with
Bayesian backfitting, collapsed trans-dimensional updates, and
residual-guided proposals. These computational strategies are designed to
improve posterior exploration across basis configurations of varying
dimensions and structures. We also establish approximation bounds for functions with
low-order interactions and heterogeneous smoothness.
These bounds relate approximation accuracy to the number
of basis terms.
Simulation studies show that MLABS achieves competitive predictive
performance across a wide range of regression settings, including
high-dimensional problems. MLABS also demonstrates robust predictive
performance across benchmark real datasets. In an environmental-mixture
application, it attains predictive accuracy comparable to Bayesian kernel
machine regression at substantially lower computational cost.

\end{abstract}

\newpage

\section{Introduction}
\label{sec:intro}

Suppose we observe $n$ pairs $\{(\mathbf{x}_i, Y_i)\}_{i=1}^n$ satisfying
\begin{equation*}\label{eq:model}
  Y_i = f(\mathbf{x}_i) + \varepsilon_i,\quad
  \varepsilon_i \overset{\mathrm{iid}}{\sim} \mathcal{N}(0,\sigma^2)
\end{equation*}
for $i = 1, \ldots, n$, where $\mathbf{x}_i \in \mathcal{X} \subseteq \mathbb{R}^p$ and
$f : \mathcal{X} \to \mathbb{R}$ is an unknown regression function.
A central challenge in multivariate nonparametric regression is to construct
models that can simultaneously accommodate nonlinear effects, interactions
among predictors, and spatially heterogeneous smoothness, while remaining
computationally feasible in moderate to high dimensions.

A standard approach represents $f$ through a basis expansion,
$
f(\mathbf{x}) \approx \sum_j g(\mathbf{x};\boldsymbol{\theta}_j)\beta_j,
$
where $g(\cdot;\boldsymbol{\theta}_j)$ denotes a basis function indexed by
structural parameters $\boldsymbol{\theta}_j$, and $\beta_j$ is the
associated coefficient. The flexibility of such a model depends on both
the basis system, which determines the features that can be represented,
and the mechanism that controls the size and structure of the expansion.
The quality of a method therefore depends not only on the expressive power
of its basis functions, but also on whether the induced model search problem
can be handled effectively. Existing approaches make different trade-offs
among representation, adaptation, and computation. We briefly review three broad classes of
methods that illustrate these trade-offs.

\paragraph{Kernel-based methods.}
Kernel methods provide a flexible framework for nonlinear regression through
kernel expansions and reproducing kernel Hilbert spaces
\citep{kimeldorf1971some, wahba1990spline}.
In Bayesian nonparametrics, the L\'evy adaptive regression kernels (LARK)
model \citep{tu2006bayesian} places a L\'evy process prior on a kernel
expansion to obtain sparse adaptive representations, with multivariate
extensions developed by \citet{ouyang2008bayesian}.
Gaussian process regression provides a closely related Bayesian framework;
Bayesian kernel machine regression \citep{bobb2015bayesian}, for example,
has been widely used for environmental-mixture analysis.
Although kernel methods are highly flexible, commonly used kernel families
often impose a common smoothness structure through their covariance kernels
or bandwidth parameters. This can make it difficult to adapt to spatially
varying smoothness, sharp transitions, and localized irregular features. Moreover, repeated computation involving
dense kernel matrices and the selection of kernel parameters can become
computationally expensive for large samples and high-dimensional spaces.

\paragraph{Spline-based methods.}
Spline methods provide local support and strong approximation properties.
These properties make them natural tools for adaptive nonparametric
regression. Multivariate adaptive regression splines
\citep[MARS;][]{friedman1991multivariate} represent nonlinear effects and
interactions through products of hinge functions. Bayesian extensions
provide regularization and uncertainty quantification
\citep{denison1998bayesian,francom2018sensitivity}.
In the univariate setting, \citet{park2023levy} proposed L\'evy adaptive
B-spline regression (LABS), which places a L\'evy process prior on a
B-spline expansion and learns the number of basis functions and their knot
locations from the data. Spline representations have also appeared in recent neural-network models.
The Kolmogorov--Arnold network (KAN) \citep{liu2024kan}, for example, uses
learnable spline functions within a neural architecture. Such approaches
illustrate the flexibility of spline-based representations, but their
structural choices, such as spline degree, knot or grid configuration, and
interaction architecture, are typically specified or tuned separately from
model fitting.

\paragraph{Ensemble-based methods.}
Tree-based ensemble methods form another major class of flexible regression
tools. Random forests \citep{breiman2001random} and boosting methods
\citep{friedman2001greedy,chen2016xgboost} combine multiple trees to improve
predictive performance. In the Bayesian setting, Bayesian additive regression
trees \citep[BART;][]{chipman2010bart} represent the regression function as a
sum of decision trees. Subsequent extensions incorporate sparsity-inducing
priors for variable selection \citep{linero2018dart} and soft decision-tree
formulations \citep{linero2018sbart}. These methods often achieve strong predictive accuracy and capture
nonlinearities and interactions effectively through recursive partitioning
of the predictor space. However, tree-based ensembles typically use
piecewise-constant or piecewise-smooth representations. As a result, they
provide less direct control over the local degree of smoothness than spline
bases and do not explicitly represent interactions through localized smooth
basis functions.

\bigskip
These considerations motivate the present work. An effective method for
multivariate nonparametric regression should capture nonlinear effects and
interactions, adapt to heterogeneous smoothness, determine model complexity
from the data, and remain computationally practical as the dimension
increases. We propose the \textbf{Multivariate L\'evy Adaptive B-Spline
Regression (MLABS)} model, a multivariate extension of the univariate LABS
framework \citep{park2023levy}. MLABS represents the regression function as a
random sum of low-order tensor-product B-spline basis functions under a finite
L\'evy random measure prior. The L\'evy prior induces a sparse random basis
expansion, and the number of basis terms, active predictors, interaction
orders, spline degrees, and knot locations are learned jointly from the data.
The local support and degree flexibility of the B-spline factors accommodate
both smooth global trends and localized irregular features. The low-order
product structure explicitly represents interactions without requiring a
full tensor-product expansion over all predictors.

Because both the number and structure of the basis functions are unknown,
posterior inference must explore a trans-dimensional space of basis
configurations. We therefore construct a reversible-jump MCMC sampler
\citep{green1995reversible} tailored to the MLABS representation. The
sampler combines Bayesian backfitting \citep{hastie2000bayesian} with
collapsed birth/death moves, residual-guided proposals for active predictor
and knot selection, and joint degree--knot refresh moves. These computational
strategies are designed to improve posterior exploration across basis
configurations of varying dimensions and structures.

We also study the approximation properties of the MLABS function class.
We establish universal approximation when interactions among all
predictors are allowed and derive explicit approximation bounds for
functions with low-order ANOVA structure. These bounds relate
approximation accuracy to the number of basis terms and the smoothness
of each component. For a fixed maximum interaction order, a given
relative approximation accuracy can be achieved with a number of
basis terms that grows at most polynomially with the number of
predictors. This bound assumes uniform smoothness conditions
and approximation constants across components. The same construction
also permits spline degrees and knot counts to vary across components.
These results provide theoretical support for the low-order
tensor-product representation used in MLABS.

The main contributions of this paper are fourfold.
First, we introduce MLABS, a Bayesian nonparametric regression model that
combines a sparse L\'evy random measure prior with low-order tensor-product
B-spline basis functions to adapt to nonlinear effects, interactions, and
heterogeneous local structure.
Second, we construct a reversible-jump MCMC sampler tailored to the
trans-dimensional MLABS representation, with collapsed structural updates and
residual-guided proposals for posterior exploration.
Third, we establish universal approximation and derive
approximation bounds for functions with low-order ANOVA
structure. These bounds relate approximation accuracy to
the number of basis terms and characterize the dependence
on the number of predictors.
Finally, numerical studies demonstrate competitive and robust predictive
performance across a broad range of simulated and real-data settings, including
high-dimensional regression and environmental-mixture analysis.

The remainder of the paper is organized as follows.
Section~\ref{sec:background} reviews the LARK and LABS models.
Section~\ref{sec:model} introduces the MLABS model and
Section~\ref{sec:computation} describes posterior computation.
Section~\ref{sec:theory} studies approximation-theoretic properties of the
MLABS function class.
Section~\ref{sec:simulations} presents simulation results, and
Section~\ref{sec:applications} reports real-data applications.
Section~\ref{sec:discussion} concludes with a discussion.

\section{Background}
\label{sec:background}

This section reviews L\'evy adaptive regression kernels and L\'evy adaptive
B-spline regression. We first describe how a finite L\'evy random measure
induces a random basis expansion and then review the univariate LABS model
based on B-spline generating functions.

\subsection{L\'evy adaptive regression kernels}

Let $\Omega$ denote a parameter space for a family of generating functions
$g(\cdot,\omega)$, with $\omega\in\Omega$. A L\'evy random measure $L$ with
L\'evy measure $\nu$ can be represented through a Poisson random measure $N$
on $\mathbb{R}\times\Omega$ with mean measure $\nu$ as
\[
L(A)
=
\int\int_{\mathbb{R}\times A}
\beta\,N(d\beta,d\omega),
\qquad A\subset\Omega.
\]
For a measurable generating function
$g:\mathcal{X}\times\Omega\to\mathbb{R}$, this construction defines the
random function
\begin{align}
\begin{split}
\label{eq:PR}
f(x)
&=
\int_{\Omega} g(x,\omega)\,L(d\omega)   \\
&=
\int\int_{\mathbb{R}\times\Omega}
g(x,\omega)\beta\,N(d\beta,d\omega),
\; x\in\mathcal{X}.
\end{split}
\end{align}
The generating function determines the dictionary of candidate basis
functions, whereas the L\'evy random measure determines which basis
functions are included and their associated coefficients.

When the total mass $\nu(\mathbb{R}\times\Omega)$ is finite, $L$ is a
compound Poisson random measure and can be written as
\[
L(d\omega)
=
\sum_{j=1}^{J}
\beta_j\delta_{\omega_j}(d\omega),
\]
where
\[
J\sim \mathrm{Poi}\{\nu(\mathbb{R}\times\Omega)\},
\quad
(\beta_j,\omega_j)
\stackrel{\mathrm{iid}}{\sim}
\frac{\nu(d\beta,d\omega)}
     {\nu(\mathbb{R}\times\Omega)}.
\]
Consequently, \eqref{eq:PR} reduces to the random finite expansion
\begin{equation*}
\label{eq:PR2}
f(x)
=
\sum_{j=1}^{J}
g(x,\omega_j)\beta_j.
\end{equation*}
Thus, a finite L\'evy random measure induces a prior jointly on the number
of basis functions, their coefficients, and their structural parameters.
Because only a finite random subset of the candidate basis family is active
in each realization, the resulting representation is naturally sparse.

The L\'evy adaptive regression kernels model
\citep[LARK;][]{tu2006bayesian,wolpert2011stochastic} uses this construction
as a Bayesian nonparametric regression prior. The generating functions are
chosen from an overcomplete family, such as Gaussian kernels, Laplace
kernels, or Haar wavelets. The L\'evy random measure selects a sparse random
collection from this family. The general LARK framework also allows
infinite-activity L\'evy measures, but the finite-measure representation is
sufficient for the development of MLABS.

\subsection{L\'evy adaptive B-spline regression}

L\'evy adaptive B-spline regression
\citep[LABS;][]{park2023levy} specializes the LARK construction by using
B-spline basis functions as the generating functions. For a univariate
predictor $x\in\mathcal{X}$ and a set $S$ of allowable spline degrees, LABS
represents the regression function as
\begin{equation}
\label{eq:meanfunc}
f(x)
=
\sum_{k\in S}
\int_{\Omega_k}
B_k(x;\boldsymbol{\xi}_k)\,
L_k(d\boldsymbol{\xi}_k),
\end{equation}
where $B_k(x;\boldsymbol{\xi}_k)$ is a degree-$k$ B-spline determined by an
ordered knot sequence
$
\boldsymbol{\xi}_k
=
(\xi_{k,1},\ldots,\xi_{k,k+2})
\in\Omega_k.
$
B-spline basis functions have local support, and their degree determines
their polynomial order and smoothness. LABS allows both spline degrees and
knot locations to vary across basis functions. This flexibility provides a
local representation for functions whose regularity changes across the
predictor domain.

Under a finite L\'evy measure for each degree $k$, the stochastic integral
in \eqref{eq:meanfunc} has the random finite-sum representation
\begin{equation*}
\label{eq:labs_finite}
f(x)
=
\beta_0+
\sum_{k\in S}
\sum_{l=1}^{J_k}
B_k(x;\boldsymbol{\xi}_{k,l})\beta_{k,l},
\end{equation*}
where $J_k$ is the random number of degree-$k$ basis functions. A convenient
hierarchical specification is
\[
J_k\mid M_k
\sim
\mathrm{Poi}(M_k),
\quad
M_k
\sim
\mathrm{Ga}(a_{\gamma_k},b_{\gamma_k}),
\]
together with
\[
\beta_{k,l}
\stackrel{\mathrm{iid}}{\sim}
N(0,\phi_k^2),
\quad
\boldsymbol{\xi}_{k,l}
\stackrel{\mathrm{iid}}{\sim}
\mathcal{U}(\Omega_k),
\quad
l=1,\ldots,J_k.
\]
The random counts $J_k$ make the dimension of the basis expansion unknown.
Posterior inference therefore requires trans-dimensional updates, which LABS
implements through reversible-jump Markov chain Monte Carlo
\citep[RJMCMC;][]{green1995reversible}.

The LABS construction is adaptive in two complementary ways. First, the
L\'evy random measures determine the number of B-spline basis functions and
their knot locations. Second, basis functions of different degrees can
coexist in the same regression fit, so different components can have
different levels of smoothness.

The original LABS formulation is univariate, so each basis function depends
on a single predictor and cannot represent interactions among predictors.
In addition, separate L\'evy random measures are assigned to different
spline degrees. The MLABS model of the next section removes both restrictions
through low-order tensor-product B-spline basis functions under a single
L\'evy random measure.
\section{Proposed Model}
\label{sec:model}

\subsection{Model Specification}
\label{subsec:model_spec}

We propose Multivariate L\'evy Adaptive B-Spline Regression (MLABS), a
multivariate extension of the univariate LABS representation in
\eqref{eq:meanfunc}.
To avoid a full tensor-product expansion over all predictors, MLABS uses
low-order tensor-product B-spline basis functions. Each basis term is a
product of a small number of univariate B-spline factors. This construction
represents nonlinear main effects and low-order interactions while controlling
the growth of model complexity. As in MARS
\citep{friedman1991multivariate}, the basis has a low-order tensor-product
structure, but MLABS replaces hinge functions with B-spline factors whose
degrees and knot sequences are inferred from the data. A comparison is given
in Section~\ref{subsec:mars_comparison}.

For the $j$th basis term, define
\begin{equation}
\label{eq:tpb}
\mathbf{B}_j(\mathbf{x}_i;\boldsymbol{\psi}_j)
:=
\prod_{l=1}^{K_j}
B_{c_l^{(j)}}\!
\left(x_{i,\nu_l^{(j)}};\boldsymbol{\xi}_l^{(j)}\right),
\end{equation}
where $K_j\in\{1,\ldots,K_{\max}\}$ is the interaction order,
$c_l^{(j)}\in S$ is the degree of the $l$th B-spline factor,
$\nu_l^{(j)}\in\{1,\ldots,p\}$ is the active predictor index, and
$\boldsymbol{\xi}_l^{(j)}$ is the corresponding knot sequence. Thus,
$K_j=1$ represents a nonlinear main effect, whereas $K_j\geq2$ represents
an interaction among distinct predictors.

For a predictor index $\nu$ and degree $c$, let $\mathcal{X}_{\nu}^{(c + 2)}$ be a collection of knots $(\xi_1, \ldots, \xi_{c + 2})$ satisfying $a_\nu < \xi_1 < \cdots < \xi_{c + 2} < b_\nu$.
An interval $[a_\nu,b_\nu]$ is the knot domain for the $\nu$th predictor,
chosen as a slightly expanded interval containing its observed range. The
single-factor parameter space is
\[
\Lambda
=
\bigsqcup_{c\in S}
\bigsqcup_{\nu=1}^{p}
\{c\}\times \{\nu\}\times \mathcal{X}_{\nu}^{(c+2)}.
\]
Thus, each element of $\Lambda$ specifies a degree, an active predictor, and
an ordered knot sequence for one univariate B-spline factor. We write
$\boldsymbol{\psi}_j := (K_j,\boldsymbol{\omega}_j),$ where
\[
\boldsymbol{\omega}_j
:=
\left(
(c_l^{(j)},\nu_l^{(j)},\boldsymbol{\xi}_l^{(j)})
\right)_{l=1}^{K_j}
\in \Lambda^{K_j},
\]
and define the structural parameter space for the tensor-product basis
functions as
\[
\Omega'
=
\bigsqcup_{K=1}^{K_{\max}}
\{K\}\times \Lambda^K .
\]

The MLABS regression function is
\begin{equation*}
\label{eq:mofm}
f(\mathbf{x}_i)
=
\beta_0
+
\sum_{j=1}^{J}
\mathbf{B}_j(\mathbf{x}_i;\boldsymbol{\psi}_j)\beta_j,
\end{equation*}
where $\beta_0$ is an intercept, $J$ is the random number of basis terms,
$\boldsymbol{\psi}_j$ denotes the structural parameters of the $j$th basis
term, and $\beta_j$ is its coefficient. Let
$\bar{J}_M := M e^{-\lambda_J}$ denote the expected number of basis terms. 
The hierarchical model is
\begin{equation}
\label{eq:mlabs}
\begin{gathered}
Y_i \mid \mathbf{x}_i
\stackrel{\mathrm{ind}}{\sim}
\mathcal{N}\!\left(f(\mathbf{x}_i),\sigma^2\right),
\\
f(\mathbf{x}_i)
=
\beta_0
+
\sum_{j=1}^{J}
\mathbf{B}_j(\mathbf{x}_i;\boldsymbol{\psi}_j)\beta_j,
\\[5pt]
\sigma^2
\sim
\mathrm{IG}\!\left(\frac{r}{2},\frac{rR}{2}\right),
\\
M
\sim
\mathrm{Ga}(a_\gamma,b_\gamma),
\\
J\mid M,\lambda_J
\sim
\mathrm{Poi}\!\left(\bar{J}_M\right),
\\[5pt]
\beta_j\mid M
\stackrel{\mathrm{iid}}{\sim}
\mathcal{N}(0,\phi_M^2),
\quad
\phi_M
=
c_{\mathrm{shrink}}
\frac{\mathrm{sd}(\mathbf{Y})}{\sqrt{\bar{J}_M}},
\\[5pt]
\boldsymbol{\pi}
\sim
\mathrm{Dir}(\alpha/p,\ldots,\alpha/p),
\\
K_j
\stackrel{\mathrm{iid}}{\sim}
\mathrm{Categorical}(w_1,\ldots,w_{K_{\max}}),
\\[5pt]
\boldsymbol{\nu}^{(j)}
\mid K_j,\boldsymbol{\pi}
\sim
\mathrm{WOR}_{K_j}(\boldsymbol{\pi}),
\qquad
c_l^{(j)}
\stackrel{\mathrm{iid}}{\sim}
\pi_d,
\\[5pt]
\boldsymbol{\xi}_l^{(j)}
\mid
\nu_l^{(j)},c_l^{(j)}
\stackrel{\mathrm{ind}}{\sim}
\pi_\xi\!\left(\cdot \mid \nu_l^{(j)},c_l^{(j)}\right),
\end{gathered}
\end{equation}
for $j=1,\ldots,J$ and $l=1,\ldots,K_j$. The priors on the interaction
order $K_j$, the active predictors $\boldsymbol{\nu}^{(j)}$, the degrees
$c_l^{(j)}$, and the knot sequences $\boldsymbol{\xi}_l^{(j)}$ are specified
below. Here $\mathrm{WOR}_{K_j}(\boldsymbol{\pi})$ denotes weighted sampling
without replacement of $K_j$ active predictors from $\{1,\ldots,p\}$ with
weights $\boldsymbol{\pi}$. Nonlinear effects of a single predictor, such as
quadratic or threshold-like patterns, can be represented by sums of several
$K_j=1$ basis functions with different degrees and knot locations.

The prior on the interaction order determines the relative preference for
main effects and interaction terms. For example, when $K_{\max}=2$,
$(w_1,w_2)=(0.8,0.2)$ favors main effects but still assigns positive
probability to two-way interactions, whereas $(w_1,w_2)=(0,1)$ restricts
all basis functions to two-way interactions. In practice, we use small values
of $K_{\max}$, typically $K_{\max}\leq3$, because high-order interactions
increase the dimension of the knot space and the computational cost of
trans-dimensional updates.

The variable-selection probabilities
$\boldsymbol{\pi}=(\pi_1,\ldots,\pi_p)$ encourage sparse use of predictors.
Conditional on $K_j$ and $\boldsymbol{\pi}$, the active predictor vector
$\boldsymbol{\nu}^{(j)}
=(\nu_1^{(j)},\ldots,\nu_{K_j}^{(j)})$
is sampled sequentially without replacement. For an ordered vector
$(\ell_1,\ldots,\ell_{K_j})$ of distinct indices,
\begin{align*}
&p_{\mathrm{WOR}}
\left(
\boldsymbol{\nu}^{(j)}
=
(\ell_1,\ldots,\ell_{K_j})
\mid
K_j,\boldsymbol{\pi}
\right) \\
&=
\prod_{m=1}^{K_j}
\frac{\pi_{\ell_m}}
     {1-\sum_{h=1}^{m-1}\pi_{\ell_h}}.
\end{align*}

This prior prevents the same predictor from appearing more than once within
a single tensor-product basis, so $K_j$ has a clear interpretation as the
interaction order.

For the degree of each B-spline factor, we place a uniform prior on the
allowed set $S$,
\begin{equation*}
\label{eq:degree_prior}
\pi_d(c)=\frac{1}{|S|},
\qquad
c\in S.
\end{equation*}
If $S$ is a singleton, the degree is fixed. Given the active predictor
$\nu_l^{(j)}$ and degree $c_l^{(j)}$, the knot sequence
\[
\boldsymbol{\xi}_l^{(j)}
=
\left(
\xi_{l,1}^{(j)},\ldots,\xi_{l,c_l^{(j)}+2}^{(j)}
\right)
\]
lies within the knot domain of the corresponding predictor and is ordered
by construction. The knot prior $\pi_\xi$ depends on the spline degree. For even degrees,
it is uniform on the ordered-knot domain. For odd degrees, the central
knot is drawn from the density
\begin{equation}
\label{eq:odd_center_prior_main}
h_\ell(u)
=
(1-\eta_0)H_\ell(u)
+
\eta_0
\frac{\mathbf1\{a_\ell<u<b_\ell\}}{b_\ell-a_\ell},
\end{equation}
where $\ell$ denotes the active predictor and $H_\ell$ is a histogram
density on its observed range: each bin receives mass equal to the
proportion of observations it contains, distributed uniformly within
the bin. We set $H_\ell=0$ outside this range and fix the uniform
mixture weight at $\eta_0=0.05$, ensuring that $h_\ell$ is positive
throughout the knot domain. This prior depends only on the predictor
design, not on the responses or residuals.

Conditional on the central knot, the remaining knots are generated
from ordered-uniform distributions on its left and right. The full
knot vector therefore has a joint density with respect to Lebesgue
measure on the ordered-knot domain. The bin construction and full
knot-prior densities are provided in Appendix~\ref{sec:supp_comp}.

The representation in \eqref{eq:mlabs} also has a finite
L\'evy-random-measure interpretation. Conditional on $M$,
$\boldsymbol\pi$, and $\mathbf x$, let $L$ be a finite
L\'evy random measure on $\Omega'$ with intensity measure
\begin{equation*}
\label{eq:levy_intensity}
\nu_M(d\beta,d\boldsymbol{\psi})
=
\bar{J}_M\,
\mathcal{N}(d\beta;0,\phi_M^2)\,
\pi(d\boldsymbol{\psi}),
\end{equation*}
where $\pi(d\boldsymbol\psi)$ denotes the structural probability
measure conditional on $\boldsymbol\pi$ and $\mathbf x$, induced by the priors on
$K_j$, $\boldsymbol{\nu}^{(j)}$, $\{c_l^{(j)}\}$, and
$\{\boldsymbol{\xi}_l^{(j)}\}$. Since
$\nu_M(\mathbb{R}\times\Omega')=\bar{J}_M<\infty$, $L$ is a
compound-Poisson random measure and can be written as
\[
L(d\boldsymbol{\psi})
=
\sum_{j=1}^{J}
\beta_j\delta_{\boldsymbol{\psi}_j}(d\boldsymbol{\psi}),
\]
with
\begin{align*}
J\mid M
&\sim
\mathrm{Poi}(\bar{J}_M),
\\
\beta_j\mid M
&\stackrel{\mathrm{iid}}{\sim}
\mathcal{N}(0,\phi_M^2),
\\
\boldsymbol{\psi}_j
\mid \boldsymbol\pi,\mathbf x
&\stackrel{\mathrm{iid}}{\sim}
\pi,
\end{align*}
where $\beta_j$ and $\boldsymbol{\psi}_j$ are independent conditional
on $M$, $\boldsymbol\pi$, and $\mathbf x$. It follows that
\begin{align*}
f(\mathbf{x})\mid M,\boldsymbol\pi,\mathbf x
&=
\beta_0
+
\int_{\Omega'}
\mathbf{B}(\mathbf{x};\boldsymbol{\psi})
L(d\boldsymbol{\psi})
\\
&=
\beta_0
+
\sum_{j=1}^{J}
\mathbf{B}_j(\mathbf{x};\boldsymbol{\psi}_j)\beta_j.
\end{align*}
Hence, conditional on $M$, $\boldsymbol\pi$, and $\mathbf x$, a single finite L\'evy random measure on the joint
space $\Omega'$ governs both the number of basis functions and their
structural parameters.

This unified construction differs from the original LABS model, which
assigns a separate L\'evy random measure to each spline degree. In MLABS,
the degree of each B-spline factor is part of the joint structural parameter,
together with the interaction order, active predictors, and knot locations.
These structural components can therefore be updated jointly in the
trans-dimensional sampler described in Section~\ref{sec:computation}.

The prior scale in \eqref{eq:mlabs} follows a standard shrinkage principle
for additive regression models \citep{chipman2010bart}. As the expected
number of additive terms increases, the prior scale of each individual
coefficient decreases. Since $\phi_M$ depends on $\bar{J}_M$ rather than on
the realized value of $J$, the coefficient prior remains well defined under
the compound-Poisson representation. For fixed $\mathbf{x}$,
\begin{align*}
\mathrm{Var}\!\left(
f(\mathbf{x})\mid M,\boldsymbol\pi,\mathbf x
\right)
&=
\bar{J}_M \phi_M^2
\mathbb{E}_{\pi}
\left[
\mathbf{B}(\mathbf{x};\boldsymbol{\psi})^2
\right]
\\
&=
c_{\mathrm{shrink}}^2
\mathrm{sd}(\mathbf{Y})^2
\mathbb{E}_{\pi}
\left[
\mathbf{B}(\mathbf{x};\boldsymbol{\psi})^2
\right],
\end{align*}
so the prior signal variance does not increase with the expected number of
basis terms.

In our numerical studies, we fix $\beta_0=\bar{Y}$ and use small values of
$K_{\max}$. The hyperparameters
$(\lambda_J,c_{\mathrm{shrink}},\{w_k\})$ control the expected number of
basis terms, the scale of individual basis effects, and the distribution of
interaction orders, respectively.

\subsection{Comparison with the MARS Basis}
\label{subsec:mars_comparison}

The MLABS and MARS bases both construct multivariate basis functions as
products of univariate factors indexed by an interaction order $K_j$ and a
set of active predictors. Their main difference lies in the form of the
univariate factors. In MARS, the $j$th basis function is
\[
N_j(\mathbf{x}_i)
=
\prod_{l=1}^{K_j}
\bigl[
s_l^{(j)}
\{x_{i,\nu_l^{(j)}}-t_l^{(j)}\}
\bigr]_+,
\]
where $s_l^{(j)}\in\{-1,+1\}$ is a sign indicator,
$t_l^{(j)}$ is a knot, and $[a]_+=\max(a,0)$. MLABS instead uses
B-spline factors whose degrees and knot sequences are inferred from the data.

To illustrate the resulting difference in basis geometry, we consider the
two-dimensional discontinuous test function used in
Section~\ref{sec:simulations}. For this illustration, we set
$K_{\max}=2$ and
\[
\Pr(K_j=1)=0,
\qquad
\Pr(K_j=2)=1,
\]
so every MLABS basis contains both $x_1$ and $x_2$. We compare the resulting
basis representation with Bayesian MARS \citep{francom2018sensitivity}.

\begin{figure*}[ht!]
\centering
\includegraphics[width=\textwidth]{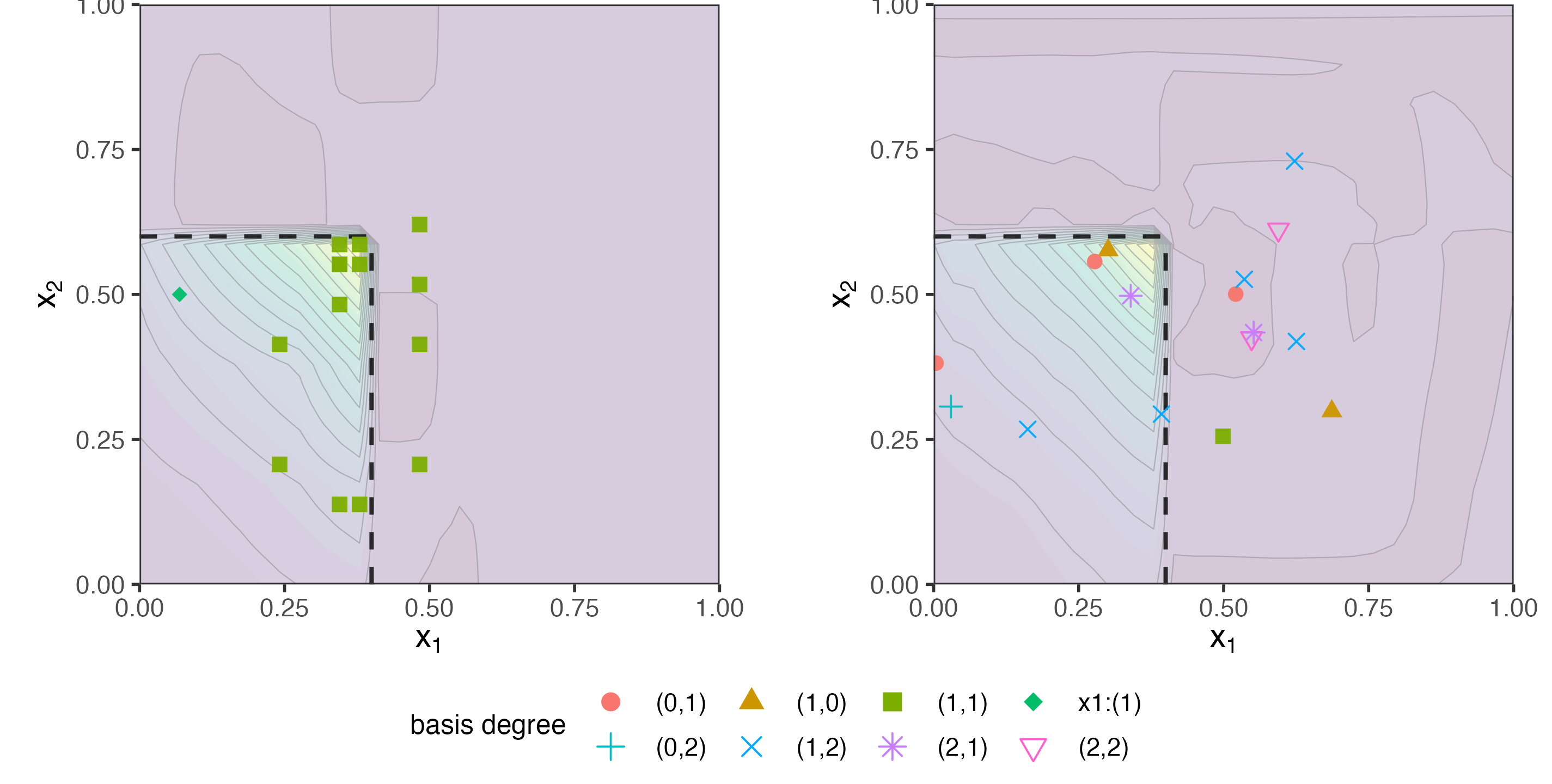}
\caption{%
Knot placement under Bayesian MARS (left) and MLABS (right) for the
two-dimensional discontinuous test function. Filled contours show the
fitted mean surfaces, and the dashed L-shaped line marks the true interior
discontinuity boundary
$\left\{
(x_1,x_2):
x_1=0.4,\; 0\leq x_2\leq 0.6
\right\}
\cup
\left\{
(x_1,x_2):
0\leq x_1\leq 0.4,\; x_2=0.6
\right\}.
$
Symbols indicate the degrees of the corresponding univariate factors.}
\label{fig:knot_compare}
\end{figure*}

\begin{figure*}[t!]
\centering

\begin{minipage}{\textwidth}
    \centering
    \includegraphics[width=0.8\textwidth]{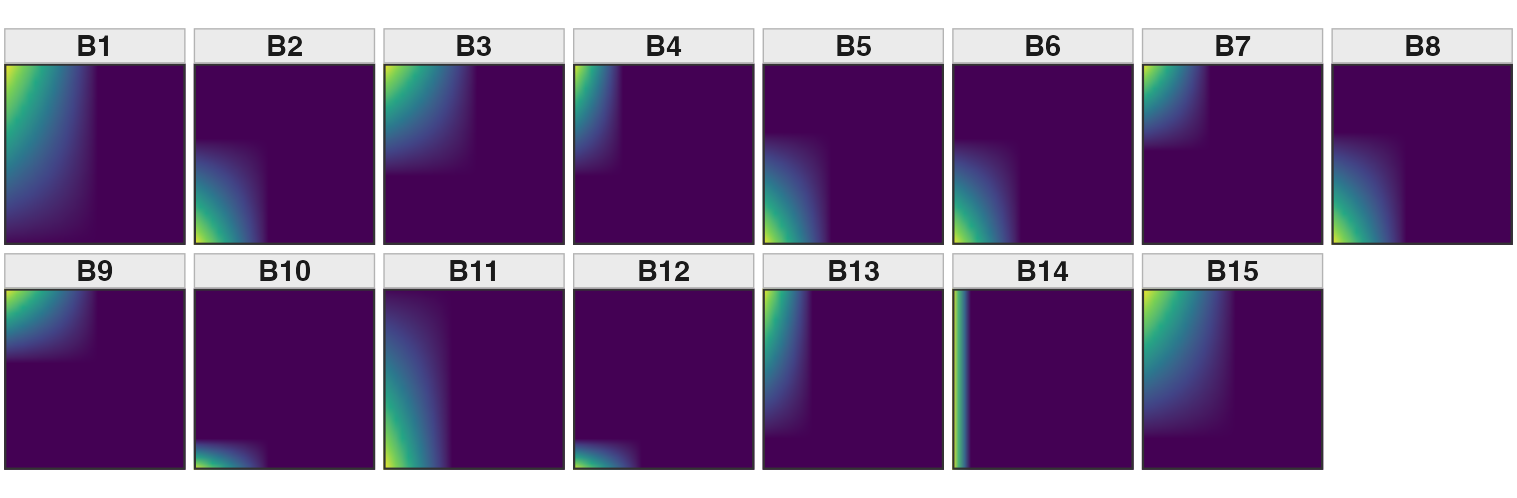}
    \par\smallskip
    {\small (a) Bayesian MARS: products of piecewise-linear
    hinge functions.}
\end{minipage}

\vspace{0.4cm}

\begin{minipage}{\textwidth}
    \centering
    \includegraphics[width=0.8\textwidth]{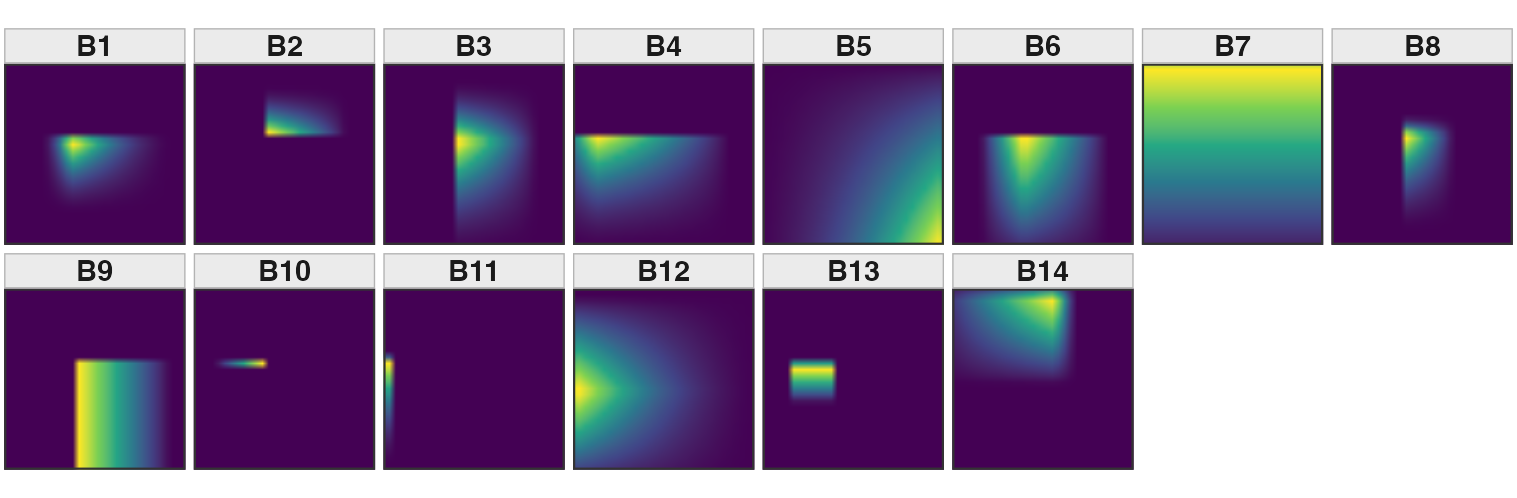}
    \par\smallskip
    {\small (b) MLABS: tensor-product B-spline basis functions.}
\end{minipage}

\caption{
Representative basis functions from the Bayesian MARS and
MLABS fits to the two-dimensional Genz surface. Each small
panel displays one product basis over the $(x_1,x_2)$ domain,
with brighter values indicating larger basis-function values.
}
\label{fig:basis_compare}
\end{figure*}

\paragraph{Knot placement.}
Each univariate MARS factor uses a single knot location together with a hinge direction, whereas a degree-$c_l^{(j)}$ MLABS factor is associated with an ordered sequence of $c_l^{(j)}+2$ knots. For MLABS, this local knot sequence jointly determines the location, support, and shape of the corresponding B-spline factor.

Figure~\ref{fig:knot_compare} illustrates this distinction. In the Bayesian MARS fit, many knot locations are concentrated near the discontinuity boundary, especially around the corner $(0.4,0.6)$. Because MARS represents changes through continuous piecewise-linear hinges, several hinge-product terms with nearby knots are used to approximate the abrupt transition. In contrast, MLABS employs locally supported B-spline factors and can include degree-zero factors capable of representing jump-like behavior. In the fitted MLABS model, the knot locations are more dispersed: some occur near the discontinuity boundary, whereas others lie within the nonzero region, consistent with representing the smooth variation inside that region. Thus, a MARS knot specifies the change point of an individual hinge factor, whereas an MLABS knot sequence jointly determines the support and shape of a local B-spline factor.

\paragraph{Degree and support flexibility.}
MARS uses piecewise-linear hinge functions as its univariate
building blocks. Each MLABS factor has its own degree
$c_l^{(j)}\in S$ and knot sequence, so MLABS basis functions
can vary substantially in support and shape.

Figure~\ref{fig:basis_compare} illustrates these differences through representative basis functions from the Bayesian MARS and MLABS fits to the two-dimensional discontinuous test function. Bayesian MARS basis functions are products of directional hinge functions, so their nonzero regions typically extend from their knot locations toward the boundaries of the predictor domain. In contrast, the support of an MLABS basis function is controlled by the knot sequences of its B-spline factors. Consequently, the fitted MLABS bases exhibit a wide range of geometries: some have broad support and describe gradual variation, whereas others are highly localized to small regions of the predictor space. This combination of broad and localized basis functions allows a single MLABS expansion to accommodate both smooth variation and abrupt local features.

\section{Posterior Computation}
\label{sec:computation}

Since both the number and structure of the basis functions are unknown,
posterior inference for MLABS requires trans-dimensional updates. We
construct a reversible-jump MCMC sampler \citep{green1995reversible} that
combines Bayesian backfitting with collapsed structural moves. For birth and degree--knot refresh moves, we integrate out the
coefficient of the candidate basis and evaluate its structure through
a closed-form marginal likelihood. We also use
the current residuals to guide active-predictor selection and knot
placement. Existing methods propose new knots near existing knots or use the current
model structure to guide the selection of new basis functions
\citep{dimatteo2001bayesian,nott2005efficient}. In MLABS, we use the
current residuals to guide both active-predictor selection and
B-spline knot placement within tensor-product basis functions.

\subsection{Sampler overview}
\label{subsec:overview}

Let
\[
\Theta_J
=
\{(\beta_j,\boldsymbol{\psi}_j):j=1,\ldots,J\},
\qquad
\boldsymbol{\psi}_j
=
(K_j,\boldsymbol{\omega}_j),
\]
where $\boldsymbol{\psi}_j$ contains the interaction order, active predictors,
spline degrees, and knot sequences of the $j$th basis function. Each MCMC
sweep consists of four types of updates:
\begin{enumerate}
    \item backfitting updates of the coefficients and local knot locations;
    \item joint degree--knot refresh moves for existing basis functions;
    \item collapsed birth/death moves for the number of basis functions;
    \item updates of the global parameters
    $\sigma^2$, $M$, and $\boldsymbol{\pi}$.
\end{enumerate}
All residual-guided proposals are corrected by the corresponding
Metropolis--Hastings ratios. Thus, residual information changes the proposal
mechanism but not the target posterior distribution.

\subsection{Collapsed structural score}
\label{subsec:collapsed}

Structural moves either introduce a new basis function or modify the
structure of an existing one. For a candidate basis function $\mathbf B$, define its basis vector by
\[
\boldsymbol b
=
\left(
\mathbf B(\mathbf x_1;\boldsymbol\psi),
\ldots,
\mathbf B(\mathbf x_n;\boldsymbol\psi)
\right)^\top.
\]
Let $\boldsymbol r$ denote the residual to be explained by this
candidate basis vector. Under
\[
\boldsymbol r\mid\beta,\boldsymbol b
\sim
N_n(\beta\boldsymbol b,\sigma^2\mathbf I_n),
\qquad
\beta\sim N(0,\phi_M^2),
\]
the conditional posterior distribution of the corresponding
coefficient is
\[
\beta\mid\boldsymbol r,\boldsymbol b
\sim
N(m,v^2),
\]
where
\begin{equation}
\label{eq:collapsed_moments}
v^2
=
\left(
\frac{\boldsymbol b^\top\boldsymbol b}{\sigma^2}
+
\frac{1}{\phi_M^2}
\right)^{-1},
\qquad
m
=
v^2
\frac{\boldsymbol b^\top\boldsymbol r}{\sigma^2}.
\end{equation}
After we integrate out $\beta$, the log marginal likelihood ratio
between the model with the candidate basis and the model without it is
\begin{equation}
\label{eq:collapsed_score}
\ell(\boldsymbol b;\boldsymbol r)
=
\frac{1}{2}
\log\left(\frac{v^2}{\phi_M^2}\right)
+
\frac{1}{2}\frac{m^2}{v^2}.
\end{equation}
This score evaluates candidate bases in birth and degree--knot
refresh moves. Because $\phi_M$ depends on $M$ rather
than on the realized value of $J$, the same coefficient prior scale is
used across birth, death, and within-model structural moves.

\subsection{Structural moves}
\label{subsec:structural}

\paragraph{Joint degree--knot refresh.}
For an existing basis function $j$, the spline degrees and knot
sequences are proposed jointly,
\[
\boldsymbol c^{(j)}
\rightarrow
\boldsymbol c^{(j)\prime},
\qquad
\boldsymbol\xi^{(j)}
\rightarrow
\boldsymbol\xi^{(j)\prime},
\]
while the interaction order and active predictors are held fixed.
Joint updates are used because changing a spline degree changes both
the shape of the basis and the dimension of its knot sequence.

Define the partial residual without basis function $j$ as
\[
\boldsymbol r^{(j)}
=
\boldsymbol Y
-
\beta_0\mathbf 1_n
-
\sum_{h\ne j}
\beta_h\boldsymbol b_h.
\]
Each proposed degree is drawn uniformly from the allowed set $S$.
The knots are then drawn from the proposal for that degree and the
partial residual $\boldsymbol r^{(j)}$. Let $q_{\mathrm{ref}}$
denote this joint degree--knot proposal density, and let
$\boldsymbol b_j$ and $\boldsymbol b_j'$ denote the current and
proposed basis vectors. The move is accepted with probability
$\min\{1,A_{\mathrm{ref}}\}$, where
\begin{equation}
\label{eq:refresh_main}
\begin{aligned}
\log A_{\mathrm{ref}}
=&\;
\ell(\boldsymbol b_j';\boldsymbol r^{(j)})
-
\ell(\boldsymbol b_j;\boldsymbol r^{(j)})
+
\log
\frac{
\pi_\xi(\boldsymbol\xi^{(j)\prime}\mid
\boldsymbol c^{(j)\prime},\boldsymbol\nu^{(j)})
}{
\pi_\xi(\boldsymbol\xi^{(j)}\mid
\boldsymbol c^{(j)},\boldsymbol\nu^{(j)})
}
+
\log
\frac{
q_{\mathrm{ref}}(
\boldsymbol c^{(j)},\boldsymbol\xi^{(j)}
\mid
\boldsymbol c^{(j)\prime},\boldsymbol\xi^{(j)\prime},
\boldsymbol r^{(j)})
}{
q_{\mathrm{ref}}(
\boldsymbol c^{(j)\prime},\boldsymbol\xi^{(j)\prime}
\mid
\boldsymbol c^{(j)},\boldsymbol\xi^{(j)},
\boldsymbol r^{(j)})
}.
\end{aligned}
\end{equation}
The degree-prior and degree-proposal factors cancel because both
are uniform on $S$. The knot-prior and knot-proposal terms are
evaluated under the current and proposed spline degrees. Both
proposal densities use the same partial residual
$\boldsymbol r^{(j)}$, which does not depend on the structure or
coefficient of basis $j$. If the move is accepted, the corresponding
coefficient is resampled from $N(m,v^2)$, with moments from
\eqref{eq:collapsed_moments} for the proposed basis.

\paragraph{Residual-guided birth proposal.}
Let
\[
\boldsymbol{r}
=
\boldsymbol Y-\beta_0\mathbf1_n
-\sum_{j=1}^{J}\beta_j\boldsymbol b_j
\]
denote the current residual. A birth move proposes the structural parameter
\[
\boldsymbol{\psi}_{J+1}
=
(K_{J+1},
 \boldsymbol{\nu}^{(J+1)},
 \boldsymbol{c}^{(J+1)},
 \boldsymbol{\xi}^{(J+1)})
\]
of a new basis function. Its coefficient is integrated out through
\eqref{eq:collapsed_score}.

To guide active-predictor selection, we partition the observed range
of each predictor into
\[
B
=
\min\{20,\max(5,\lfloor\sqrt{n}\rfloor)\}
\]
equal-width bins. Let $\mathcal I_{\ell b}$ denote the set of
observation indices in bin $b$ for predictor $\ell$, and let
$\bar r_{\ell b}$ denote their mean residual. Write
$\bar r=n^{-1}\sum_{i=1}^n r_i$ and define
\begin{align*}
SS_{\mathrm{within},\ell}
&=
\sum_{b:\,|\mathcal{I}_{\ell b}|>0}
\sum_{i\in\mathcal{I}_{\ell b}}
(r_i-\bar r_{\ell b})^2,
\\
SS_{\mathrm{total}}
&=
\sum_{i=1}^{n}(r_i-\bar r)^2.
\end{align*}
The residual relevance score is
\[
\rho_\ell
=
\max\left\{
0,\,
1-
\frac{SS_{\mathrm{within},\ell}}
     {SS_{\mathrm{total}}}
\right\},
\]
with $\rho_\ell=0$ when $SS_{\mathrm{total}}=0$. A large value of
$\rho_\ell$ indicates systematic residual variation along predictor
$\ell$. We tilt the current variable-selection probabilities according to
\begin{equation}
\label{eq:informed_var_main}
\widetilde{\pi}_\ell
\propto
\pi_\ell(1+\kappa\rho_\ell),
\qquad
\ell=1,\ldots,p,
\end{equation}
where $\kappa\geq0$ controls the strength of the residual guidance.
Conditional on $K_{J+1}$, the active predictors are sampled without
replacement according to $\widetilde{\boldsymbol{\pi}}$.

Knot locations are guided by local residual magnitudes. Let
$n_{\ell b}=|\mathcal I_{\ell b}|$ denote the number of observations
in bin $b$ for predictor $\ell$.
For each nonempty bin, define
\[
\overline{|r|}_{\ell b}
=
\frac{1}{n_{\ell b}}
\sum_{i\in\mathcal I_{\ell b}}|r_i|,
\qquad
d_{\ell b}(\boldsymbol r)
=
\left\{
\overline{|r|}_{\ell b}+\varepsilon_r
\right\}^{1/T},
\]
where $\varepsilon_r>0$ prevents zero observation weights and
$T\geq1$ controls their concentration. Let $b_\ell(i)$ be the bin
index of observation $i$. Its weight is
\begin{equation}
\label{eq:knot_weight_main}
w_i(\boldsymbol r;\ell)
=
\frac{
d_{\ell,b_\ell(i)}(\boldsymbol r)
}{
\displaystyle
\sum_{b:\,n_{\ell b}>0}
n_{\ell b}d_{\ell b}(\boldsymbol r)
}.
\end{equation}
Thus all observations within the same bin receive the same weight.

For an even degree $c$, we draw an observation index $s$ with
probability $w_s(\boldsymbol r;\ell)$ and use $x_{s\ell}$ as an
auxiliary anchor. We draw $(c+2)/2$ knots from $U(a_\ell,x_{s\ell})$
and the same number from $U(x_{s\ell},b_\ell)$, then sort the knots.
The anchor itself is not part of the knot vector. The proposal
density sums over all possible anchor indices.

For odd degrees, let $I_{\ell b}$ denote the interval corresponding
to bin $b$ for predictor $\ell$, so that
$\mathcal I_{\ell b}=\{i:x_{i\ell}\in I_{\ell b}\}$.
The observation weights are aggregated into bin probabilities,
\[
\omega_{\ell b}(\boldsymbol r)
=
\sum_{i\in\mathcal I_{\ell b}}w_i(\boldsymbol r;\ell),
\]
with $\omega_{\ell b}=0$ for empty bins. These probabilities define
the histogram density
\[
\widetilde g_\ell(u\mid\boldsymbol r)
=
\sum_b
\omega_{\ell b}(\boldsymbol r)
\frac{\mathbf1\{u\in I_{\ell b}\}}{|I_{\ell b}|},
\]
where $|I_{\ell b}|$ denotes the length of the bin interval.
For an odd degree $c$, the central position is $m_c=(c+3)/2$.
We propose the central knot from
\begin{equation}
\label{eq:odd_center_proposal_main}
g_\ell(u\mid\boldsymbol r)
=
(1-\eta)\widetilde g_\ell(u\mid\boldsymbol r)
+
\eta
\frac{\mathbf1\{a_\ell<u<b_\ell\}}{b_\ell-a_\ell},
\end{equation}
for $0< \eta \leq1$. We draw $A\sim g_\ell(\cdot\mid\boldsymbol r)$ and set
$\xi_{m_c}=A$. Conditional on $A$, we draw $(c+1)/2$ knots from
$U(a_\ell,A)$ and the same number from $U(A,b_\ell)$, then sort
the knots on each side. These conditional distributions coincide
with those in the knot prior.

The parameter $\eta$ controls the uniform component of the proposal
and is distinct from the fixed prior parameter $\eta_0$ in
\eqref{eq:odd_center_prior_main}. The density $h_\ell$ in that
equation is the prior density of the central knot. Let
$q_{c,\ell}(\boldsymbol\xi\mid\boldsymbol r)$ denote the joint
knot-proposal density for degree $c$ and predictor $\ell$.
For odd degrees, the conditional densities of the knots on either
side of $A$ cancel from the prior-to-proposal ratio. Hence
\begin{equation}
\label{eq:odd_knot_ratio_main}
\frac{
\pi_\xi(\boldsymbol\xi\mid c,\ell)
}{
q_{c,\ell}(\boldsymbol\xi\mid\boldsymbol r)
}
=
\frac{
h_\ell(\xi_{m_c})
}{
g_\ell(\xi_{m_c}\mid\boldsymbol r)
}.
\end{equation}
The same degree-specific knot proposals are used in refresh moves,
with $\boldsymbol r^{(j)}$ in place of the current residual
$\boldsymbol r$.

\paragraph{Birth and death acceptance.}
Let $\boldsymbol{b}_{\mathrm{new}}$ denote the proposed basis vector and let $q_{\mathrm{birth}}
(\boldsymbol{\psi}_{J+1}
\mid
\boldsymbol{r},\boldsymbol{\pi})$
denote the complete structural proposal density. Let $g_b(J)$ and
$g_d(J)$ denote the probabilities of selecting a birth or death move
at a state with $J$ basis functions. Under an exchangeable-label
convention, uniform insertion and reverse deletion probabilities
cancel. The birth move is accepted
with probability $\min\{1,A_{\mathrm{birth}}\}$, where
\begin{equation}
\label{eq:birth_main}
\begin{aligned}
\log A_{\mathrm{birth}}
&=
\ell(\boldsymbol{b}_{\mathrm{new}};\boldsymbol{r})
+
\log
\frac{
\pi(J+1\mid M,\lambda_J)
}{
\pi(J\mid M,\lambda_J)
}
+
\log
\frac{
\pi(\boldsymbol{\psi}_{J+1}\mid\boldsymbol{\pi})
}{
q_{\mathrm{birth}}
(\boldsymbol{\psi}_{J+1}
\mid
\boldsymbol{r},\boldsymbol{\pi})
}
+
\log
\frac{g_d(J+1)}{g_b(J)}.
\end{aligned}
\end{equation}
The four terms correspond to the collapsed marginal score, the prior ratio
for the number of basis functions, the correction for the structural
proposal, and the ratio of birth/death move probabilities. For the
thinned-Poisson prior,
\[
\frac{
\pi(J+1\mid M,\lambda_J)
}{
\pi(J\mid M,\lambda_J)
}
=
\frac{M e^{-\lambda_J}}{J+1}.
\]
The interaction order and spline degrees are proposed from their
priors, so these factors cancel from the structural prior--proposal ratio. The remaining correction accounts for the residual-guided
predictor and knot proposals. If the birth is accepted, the new
coefficient is drawn from $N(m,v^2)$, with moments from
\eqref{eq:collapsed_moments} for the proposed basis and the current
residual.

A death move removes a uniformly selected basis function $j$.
Its acceptance ratio is the reciprocal of the reverse-birth ratio,
which is evaluated at the proposed smaller model. In particular,
the reverse predictor and knot proposals use the residual
$\boldsymbol r_{-j}=\boldsymbol r+\beta_j\boldsymbol b_j$.

\subsection{Remaining updates}
\label{subsec:remaining_updates}

Conditional on the basis structure, the coefficients are updated by Gaussian backfitting, and local knot locations are updated by
Metropolis--Hastings random walks. The noise variance $\sigma^2$ and total-mass parameter $M$ have
conjugate updates. Because predictors within a basis are sampled
without replacement, the variable-selection probabilities
$\boldsymbol\pi$ are updated by a Metropolis--Hastings step with a
DART-style Dirichlet independence proposal \citep{linero2018dart}.
To avoid premature concentration on a small number of predictors,
$\boldsymbol\pi$ is held fixed at the uniform vector during early
burn-in. Full conditional distributions, proposal densities,
acceptance-ratio derivations, and the sampling algorithm are provided
in Appendix~\ref{sec:supp_comp}.

\section{Theoretical Results}
\label{sec:theory}

We study the approximation properties of the MLABS function
class. We first establish universal approximation when
interactions among all predictors are allowed. We then derive
approximation bounds for functions with low-order ANOVA
structure and allow spline degrees and resolutions to vary
across components. These results quantify the number of basis
terms required to achieve a given approximation accuracy.
All proofs are provided in Appendix~\ref{sec:supp_theory}.

Throughout this section, the predictor domain is $[0,1]^p$, and the
knot domain for each predictor $\nu$ satisfies $a_\nu<0<1<b_\nu$.
For a finite nonempty set $S$ of nonnegative integer degrees and
$K_{\max}\geq1$, define
\begin{align*}
    \mathcal F(J,K_{\max},S)    
    :=
   \bigg\{
    \beta_0+\sum_{j=1}^{J}\beta_j
    \prod_{l=1}^{K_j}
    B_{c_l^{(j)}}\!\left(x_{\nu_l^{(j)}};
    \boldsymbol\xi_l^{(j)}\right)
    : 
    \beta_0, \beta_j \in \mathbb{R}, \;
    1 \leq K_j \leq ( K_{\max} \wedge p)
    \bigg\}.
\end{align*}
The predictor indices within
each term are distinct, $c_l^{(j)}\in S$, and each local knot vector
is strictly increasing within its predictor's knot domain.
Zero coefficients are allowed, so $J$ also bounds the number of nonzero
terms. Set
\[
\mathcal F_\infty(K_{\max},S)
=\bigcup_{J\geq1}\mathcal F(J,K_{\max},S).
\]
No positive minimum knot separation or upper bound on coefficient
magnitudes is imposed. All errors are measured in Lebesgue $L^2$ on
the indicated domain. These results concern approximation by the
dictionary; they do not assert posterior contraction or automatic
selection of degrees and knots.

\subsection{Universal approximation}
\label{subsec:theory_ua}

\begin{theorem}[Universal approximation]
\label{thm:ua}
If $0\in S$ and $K_{\max}\geq p$, then
$\mathcal F_\infty(K_{\max},S)$ is dense in $L^2([0,1]^p)$.
Thus, for every $f\in L^2([0,1]^p)$ and every $\varepsilon>0$,
there exists $g\in\mathcal F_\infty(K_{\max},S)$ such that
\[
\|f-g\|_{L^2([0,1]^p)}<\varepsilon.
\]
\end{theorem}

Products of degree-zero B-splines represent rectangle indicators,
whose finite linear combinations are dense in $L^2$.
When $K_{\max}<p$, the same construction applies to each component
depending on at most $K_{\max}$ predictors. This motivates the
componentwise approximation results below.

\subsection{Approximation of low-order interactions}
\label{subsec:theory_rates}

Consider a representation
\begin{equation}
\label{eq:theory_anova}
f(\mathbf x)=\mu+\sum_{u\in\mathcal U}f_u(\mathbf x_u),
\end{equation}
where $\mathcal U\subseteq
\{u\subseteq\{1,\ldots,p\}:1\leq|u|\leq K_{\max}\}$,
$\mu\in\mathbb R$ and $f_u\in L^2([0,1]^{|u|})$.
We call this a low-order ANOVA representation without requiring
centering or uniqueness of the components. If $\mathcal U$ is empty,
the intercept represents $f$ exactly; otherwise, assume
$\mathcal U\ne\varnothing$ in what follows.

Let $B^s_{a,q}$ denote the Besov space with smoothness $s$,
integrability index $a$, and fine-scale index $q$.
Its definition and the approximation lemmas used below are given
in Appendix~\ref{sec:supp_theory}.

\begin{theorem}[Besov ANOVA approximation]
\label{thm:besov_anova}
Let $k\in S$ and suppose that $f$ has representation
\eqref{eq:theory_anova}. For each $u\in\mathcal U$, let
$d_u=|u|$ and suppose
\[
f_u\in B^{s_u}_{a_u,q_u}([0,1]^{d_u}),
\]
for
$0<s_u<k+1$ and
$ 0<a_u,q_u\leq\infty$.
Define
\[
\rho_u
=
s_u-d_u\left(\frac1{a_u}-\frac12\right)_+,
\]
where $(z)_+=\max\{z,0\}$, and assume $\rho_u>0$
for every $u\in\mathcal U$.
For every integer $N\geq1$, there exists
$g\in\mathcal F(J,K_{\max},S)$ such that
\[
J\leq\sum_{u\in\mathcal U}(N+k+1)^{d_u}
\]
and
\begin{equation}
\label{eq:anova_approx_bound}
\|f-g\|_{L^2([0,1]^p)}
\leq
C\sum_{u\in\mathcal U}
N^{-\rho_u}
\|f_u\|_{B^{s_u}_{a_u,q_u}([0,1]^{d_u})}.
\end{equation}
The constant $C$ depends on $k$, $K_{\max}$,
the Besov indices, and the quasi-uniformity constants,
but not on $N$ or the component functions.
\end{theorem}

The proof approximates each component on a quasi-uniform
tensor-product partition with $N$ interior knots per coordinate.
These spline approximations belong to the MLABS function class,
which also allows nonuniform and component-specific knot
configurations. Thus, the theorem gives an approximation bound
for MLABS through a particular choice of knots. The construction
depends on the component dimensions $d_u$ and does not require
a full $p$-dimensional tensor-product basis.

\begin{remark}
The same construction permits different spline degrees
and numbers of knots across components. For each
$u\in\mathcal U$, choose $k_u\in S$ with $s_u<k_u+1$
and an integer $N_u\geq1$. Under the remaining assumptions
of Theorem~\ref{thm:besov_anova}, there exists
$g\in\mathcal F(J,K_{\max},S)$ such that
\[
J\leq\sum_{u\in\mathcal U}(N_u+k_u+1)^{d_u}
\]
and
\[
\|f-g\|_{L^2([0,1]^p)}
\leq
C\sum_{u\in\mathcal U}N_u^{-\rho_u}
\|f_u\|_{B^{s_u}_{a_u,q_u}([0,1]^{d_u})}.
\]
These bounds follow from the componentwise approximation
lemma with degree $k_u$ and $N_u$ interior knots per
coordinate.
\end{remark}

\begin{corollary}[Dependence on the ambient dimension]
\label{cor:besov_cod}
Under the conditions of Theorem~\ref{thm:besov_anova},
suppose that $\mathcal U\neq\varnothing$ and define
\[
\rho_{\min}
=
\min_{u\in\mathcal U}\rho_u,
\qquad
\mathcal B
=
\sum_{u\in\mathcal U}
\|f_u\|_{B^{s_u}_{a_u,q_u}([0,1]^{d_u})}.
\]
For every $0<\varepsilon<1$, there exists
$g\in\mathcal F(J,K_{\max},S)$ such that
\[
\|f-g\|_{L^2([0,1]^p)}
\leq\varepsilon\mathcal B
\]
and
\[
J
=
O\!\left(
p^{K_{\max}}
\varepsilon^{-K_{\max}/\rho_{\min}}
\right).
\]
For fixed $k$ and $K_{\max}$, the implied constant
can be chosen independently of $p$ and $\varepsilon$
if the componentwise approximation constants are
uniformly bounded and $\rho_{\min}\geq\rho_*>0$
for a fixed $\rho_*$.
\end{corollary}

The normalization by $\mathcal B$ expresses the target
error relative to the sum of the component Besov norms.
Under the stated uniformity conditions, the bound grows
at most polynomially with $p$ for fixed $K_{\max}$ and
$\varepsilon$. The bound in Theorem~\ref{thm:besov_anova} accounts for the
number and dimensions of the active ANOVA components and
can therefore yield a sharper bound for sparse interaction
structures.

\section{Simulation Studies}
\label{sec:simulations}

We evaluate MLABS in three groups of simulation experiments. We first
consider four bivariate test surfaces with substantially different local
geometries: the smooth radial function of \citet{hwang1994regression}, a
localized Ricker wavelet, a Genz-type discontinuous function
\citep{genz1984testing}, and an obliquely discontinuous function
\citep{imaizumi2019deep}. We then consider the three benchmark regression
functions of \citet{friedman1991multivariate}. Finally, we use a
high-dimensional extension of Friedman~1 to assess robustness to a large
number of irrelevant predictors.

Predictive accuracy is measured by the root-mean-square error
\[
\mathrm{RMSE}
=
\left[
\frac{1}{n_{\mathrm{test}}}
\sum_{i=1}^{n_{\mathrm{test}}}
\left\{
f(\boldsymbol{x}_i^\star)
-
\widehat f(\boldsymbol{x}_i^\star)
\right\}^2
\right]^{1/2},
\]
where $\boldsymbol{x}_i^\star$ denotes a test point. For the surface and
Friedman benchmark experiments, the noise level is controlled by the root
signal-to-noise ratio
\[
\mathrm{RSNR}
=
\frac{
\operatorname{sd}\{f(\boldsymbol{x}_i)\}
}{
\sigma
},
\]
where $\sigma$ is the standard deviation of the Gaussian observation noise.

We compare MLABS with multivariate adaptive regression splines
\citep[MARS;][]{friedman1991multivariate}, Bayesian MARS
\citep[BASS;][]{francom2018sensitivity}, LARK with multivariate Gaussian
kernels \citep[BARK;][]{ouyang2008bayesian}, support vector machines (SVM),
neural networks (NN), random forests (RF), extreme gradient boosting (XGB),
Bayesian additive regression trees \citep[BART;][]{chipman2010bart},
SoftBART \citep[SBART;][]{linero2018sbart}, FlexBART
\citep{linero2025generalized}, and Kolmogorov--Arnold networks
\citep[KAN;][]{liu2024kan}. Implementation details and hyperparameter grids
for all methods are provided in Appendix~\ref{sec:supp_sim}.

Hyperparameters are selected by five-fold cross-validation. For MLABS, we
fix $b_\gamma=1$ and $r=R=0.01$, and select the remaining hyperparameters
that control the basis structure, model size, and regularization by
cross-validation. Complete tuning grids are given in Appendix~\ref{sec:supp_sim}. Each MLABS chain is run for $200{,}000$ iterations. The first
$100{,}000$ iterations are discarded as burn-in, and every tenth
post-burn-in draw is retained.

\subsection{Surface Test Functions}
\label{subsec:surface_sim}

For each surface, the training design consists of a $30\times30$ regular
grid on
\[
\mathcal X=[0,1]^2.
\]
Independent Gaussian noise is added at RSNR levels 1 and 5. Predictive RMSE
is evaluated on 2,500 independent test points drawn from $U([0,1]^2)$.
Each setting is independently replicated 100 times. The four regression
surfaces are
\begin{align*}
f^{(1)}(\mathbf{x})
&=
24.234\,r^2(0.75-r^2),
\\
f^{(2)}(\mathbf{x})
&=
\left(
1-\frac{r^2}{\tau^2}
\right)
\exp\left(
-\frac{r^2}{2\tau^2}
\right),
\;
\tau=0.12,
\\
f^{(3)}(\mathbf{x})
&=
\mathbf{1}_{R_0}(\mathbf{x})
\exp(4x_1+4x_2),
\\
f^{(4)}(\mathbf{x})
&=
\mathbf{1}_{R_1}(\mathbf{x})
(0.2+x_1^2+0.1x_2) +
\mathbf{1}_{R_2}(\mathbf{x})
\left\{
0.7+0.01|4x_1+10x_2-9|^{1.5}
\right\},
\end{align*}
where
\begin{align*}
r^2
&=
(x_1-0.5)^2+(x_2-0.5)^2, \quad 
R_0
=
\{
(x_1,x_2):
x_1\leq0.4,\;
x_2\leq0.6
\}, \\
R_1
&=
\{
(x_1,x_2):
x_2\geq-0.6x_1+0.75
\}, \quad 
R_2
=
\mathcal X\setminus R_1.
\end{align*}

The first surface is a smooth radially symmetric quartic polynomial
\citep{hwang1994regression}. The second is a localized Ricker wavelet with
a sharp central peak and a surrounding negative trough. The third is a
Genz-type surface \citep{genz1984testing} with axis-aligned jumps along
the boundary of $R_0$. Its largest jump is $e^4\approx54.6$ and occurs at
$(0.4,0.6)$. The fourth surface has an oblique discontinuity that separates
two regions with different functional forms \citep{imaizumi2019deep}.
We refer to the four functions as Radial, Mexican hat, Genz, and Oblique,
respectively. The true and representative fitted surfaces are shown in Appendix~\ref{sec:supp_sim}.

For this experiment, we include thin plate splines (TPS) as an additional
benchmark because they provide a standard smooth surface estimator in two
dimensions. Panel A of Table~\ref{tab:sim_main} reports the predictive results. MLABS
achieves the best average rank of 1.75. It has the lowest RMSE for the Genz
surface at both noise levels and the second-lowest RMSE in each of the
remaining six settings. TPS has the second-best average rank, while BARK
and BASS are particularly competitive for the smooth surfaces. KAN attains
the lowest RMSE for the Oblique surface at RSNR~$=5$. Overall, MLABS shows
consistent predictive performance across smooth, localized, and
discontinuous surface structures.

\subsection{Friedman's Test Functions}
\label{subsec:fried}

We next consider the three benchmark regression functions of
\citet{friedman1991multivariate}. For Friedman~1, we generate ten
independent $U(0,1)$ predictors and use
\[
f_1(\mathbf{x})
=
10\sin(\pi x_1x_2)
+
20(x_3-0.5)^2
+
10x_4
+
5x_5,
\]
so that only the first five predictors are relevant.

Friedman~2 and~3 use four independent predictors with
\begin{align*}
x_1 \sim U(0,100), \quad
x_2 \sim U(40\pi,560\pi),  \quad
x_3 \sim U(0,1),   \quad
x_4 \sim U(1,11)
\end{align*}
and regression functions
\begin{align*}
f_2(\mathbf{x})
&=
\left[
x_1^2
+
\left(
x_2x_3-\frac{1}{x_2x_4}
\right)^2
\right]^{1/2}, \quad
f_3(\mathbf{x})
=
\arctan
\left[
\frac{
x_2x_3-\frac{1}{x_2x_4}
}{
x_1
}
\right].
\end{align*}
For each setting, we generate 250 training observations and 1,000
independent test observations and consider RSNR levels 1 and 5. Each
setting is independently replicated 100 times.

Panel B of Table~\ref{tab:sim_main} reports the predictive results. MLABS achieves
the lowest average rank of 1.33 and ranks first or second in all six
settings. BASS is the next strongest method overall, followed by SBART.
These results show that MLABS remains competitive for multivariate
regression functions with nonlinear main effects and interactions.

\begin{sidewaystable*}[p]
\centering
\setlength{\tabcolsep}{2pt}
\renewcommand{\arraystretch}{1.0}

\caption{
Predictive RMSE for the simulation experiments.
Panel A reports results for the four surface functions, and Panel B reports
results for Friedman's test functions. Ranks are shown in parentheses, and
the lowest RMSE in each setting is in bold.
}
\label{tab:sim_main}
\vspace{5mm}
\textbf{Panel A: Surface test functions}

\vspace{1mm}

\begin{adjustbox}{max width=\linewidth, center}
\begin{tabular}{llrrrrrrrrrrrrr}
\toprule
Function & Noise & MLABS & SBART & BART & BARK & FlexBART & BASS &
RF & SVM & MARS & NN & XGB & KAN & TPS \\
\midrule

\multirow{2}{*}{Radial}
& RSNR = 5
& 0.0360 (2) & 0.0408 (4) & 0.0705 (9) & 0.0529 (6)
& 0.0697 (8) & \textbf{0.0295 (1)} & 0.1282 (11)
& 0.0391 (3) & 0.0983 (10) & 0.2046 (13)
& 0.1526 (12) & 0.0624 (7) & 0.0461 (5) \\
& RSNR = 1
& 0.1339 (2) & 0.1661 (6) & 0.2141 (9) & \textbf{0.1175 (1)}
& 0.2213 (10) & 0.1508 (4) & 0.5308 (12)
& 0.1539 (5) & 0.2051 (7) & 0.5370 (13)
& 0.3626 (11) & 0.2058 (8) & 0.1368 (3) \\

\multirow{2}{*}{Mexican hat}
& RSNR = 5
& 0.0110 (2) & 0.0402 (9) & 0.0270 (7) & \textbf{0.0075 (1)}
& 0.0260 (6) & 0.0255 (5) & 0.0387 (8)
& 0.0131 (4) & 0.1506 (12) & 0.1902 (13)
& 0.0911 (11) & 0.0436 (10) & 0.0125 (3) \\
& RSNR = 1
& 0.0346 (2) & 0.0697 (5) & 0.0726 (6) & \textbf{0.0302 (1)}
& 0.1323 (11) & 0.1047 (10) & 0.1042 (9)
& 0.0571 (4) & 0.1585 (12) & 0.1993 (13)
& 0.0884 (7) & 0.0954 (8) & 0.0542 (3) \\

\multirow{2}{*}{Genz}
& RSNR = 5
& \textbf{1.5303 (1)} & 2.0192 (8) & 1.7386 (4) & 2.1814 (9)
& 1.7227 (3) & 1.7851 (6) & 1.7071 (2)
& 2.6110 (10) & 3.6265 (12) & 4.8331 (13)
& 3.2392 (11) & 1.7527 (5) & 1.8241 (7) \\
& RSNR = 1
& \textbf{1.7298 (1)} & 2.6312 (3) & 2.4146 (2) & 2.9988 (6)
& 3.0441 (7) & 2.8631 (4) & 3.5146 (9)
& 3.6462 (10) & 3.9539 (12) & 4.9697 (13)
& 3.7301 (11) & 3.1331 (8) & 2.9835 (5) \\

\multirow{2}{*}{Oblique}
& RSNR = 5
& 0.0288 (2) & 0.0402 (8) & 0.0387 (5) & 0.0407 (9)
& 0.0399 (7) & 0.0467 (10) & 0.0394 (6)
& 0.0375 (4) & 0.0580 (12) & 0.0526 (11)
& 0.0586 (13) & \textbf{0.0284 (1)} & 0.0327 (3) \\
& RSNR = 1
& 0.0612 (2) & 0.0655 (4) & 0.0687 (7) & 0.0672 (6)
& 0.0705 (8) & 0.0730 (11) & 0.1284 (13)
& 0.0656 (5) & 0.0726 (10) & 0.0873 (12)
& 0.0725 (9) & 0.0627 (3) & \textbf{0.0581 (1)} \\

\midrule
Average rank
& &
\textbf{1.75 (1)} & 5.88 (5) & 6.13 (6) & 4.88 (3)
& 7.50 (9) & 6.38 (8) & 8.75 (10)
& 5.63 (4) & 10.88 (12) & 12.63 (13)
& 10.63 (11) & 6.25 (7) & 3.75 (2) \\
\bottomrule
\end{tabular}
\end{adjustbox}

\vspace{20mm}

\textbf{Panel B: Friedman's test functions}

\vspace{1mm}

\begin{adjustbox}{max width=\linewidth, center}
\begin{tabular}{llrrrrrrrrrrrr}
\toprule
Function & Noise & MLABS & SBART & BART & BARK & FlexBART & BASS &
RF & SVM & MARS & NN & XGB & KAN \\
\midrule

\multirow{2}{*}{Friedman 1}
& RSNR = 5
& \textbf{0.345 (1)} & 0.412 (3) & 1.056 (7) & 1.054 (6)
& 0.758 (5) & 0.383 (2) & 2.185 (11)
& 3.972 (12) & 0.599 (4) & 2.163 (10)
& 1.475 (9) & 1.419 (8) \\
& RSNR = 1
& \textbf{1.671 (1)} & 1.899 (2) & 2.163 (5) & 2.306 (7)
& 2.050 (3) & 2.155 (4) & 2.551 (9)
& 4.074 (12) & 2.341 (8) & 3.731 (11)
& 2.686 (10) & 2.261 (6) \\

\multirow{2}{*}{Friedman 2}
& RSNR = 5
& 19.06 (2) & 23.70 (4) & 43.98 (9) & 24.05 (5)
& 35.96 (6) & \textbf{17.37 (1)} & 50.54 (10)
& 77.25 (12) & 23.21 (3) & 43.23 (8)
& 51.36 (11) & 40.92 (7) \\
& RSNR = 1
& 72.35 (2) & 85.74 (4) & 129.78 (8) & 78.87 (3)
& 102.63 (5) & \textbf{53.20 (1)} & 151.25 (10)
& 185.82 (11) & 116.07 (7) & 137.11 (9)
& 200.68 (12) & 115.45 (6) \\

\multirow{2}{*}{Friedman 3}
& RSNR = 5
& \textbf{0.081 (1)} & 0.084 (3) & 0.097 (5) & 0.125 (10)
& 0.088 (4) & 0.083 (2) & 0.106 (6)
& 0.131 (11) & 0.139 (12) & 0.109 (7)
& 0.121 (9) & 0.115 (8) \\
& RSNR = 1
& \textbf{0.130 (1)} & 0.135 (2) & 0.145 (4) & 0.178 (8)
& 0.144 (3) & 0.157 (5) & 0.168 (7)
& 0.216 (12) & 0.196 (11) & 0.191 (10)
& 0.189 (9) & 0.161 (6) \\

\midrule
Average rank
& &
\textbf{1.33 (1)} & 3.00 (3) & 6.33 (5) & 6.50 (6)
& 4.33 (4) & 2.50 (2) & 8.83 (9)
& 11.67 (12) & 7.50 (8) & 9.17 (10)
& 10.00 (11) & 6.83 (7) \\
\bottomrule
\end{tabular}
\end{adjustbox}

\end{sidewaystable*}

\subsection{High-Dimensional Regression with Irrelevant Predictors}
\label{subsec:highdim}

Finally, we consider a high-dimensional extension of Friedman~1 to assess
robustness to irrelevant predictors. We follow the design of
\citet{linero2018sbart}. The number of predictors $p$ ranges from 5 to
1,000 on an approximately logarithmic grid, while the regression function
continues to depend only on the first five predictors. For each value of
$p$, we generate 250 training observations and 1,000 independent test
observations. Gaussian noise with variance
\[
\sigma^2\in\{1,10\}
\]
is added to the training responses. Each combination of $p$ and
$\sigma^2$ is independently replicated 10 times. Hyperparameters are
selected separately for each value of $p$ whenever computationally
feasible, and test RMSE is averaged over the 10 replications.

\begin{figure*}[ht!]
\centering
\includegraphics[width=\textwidth]{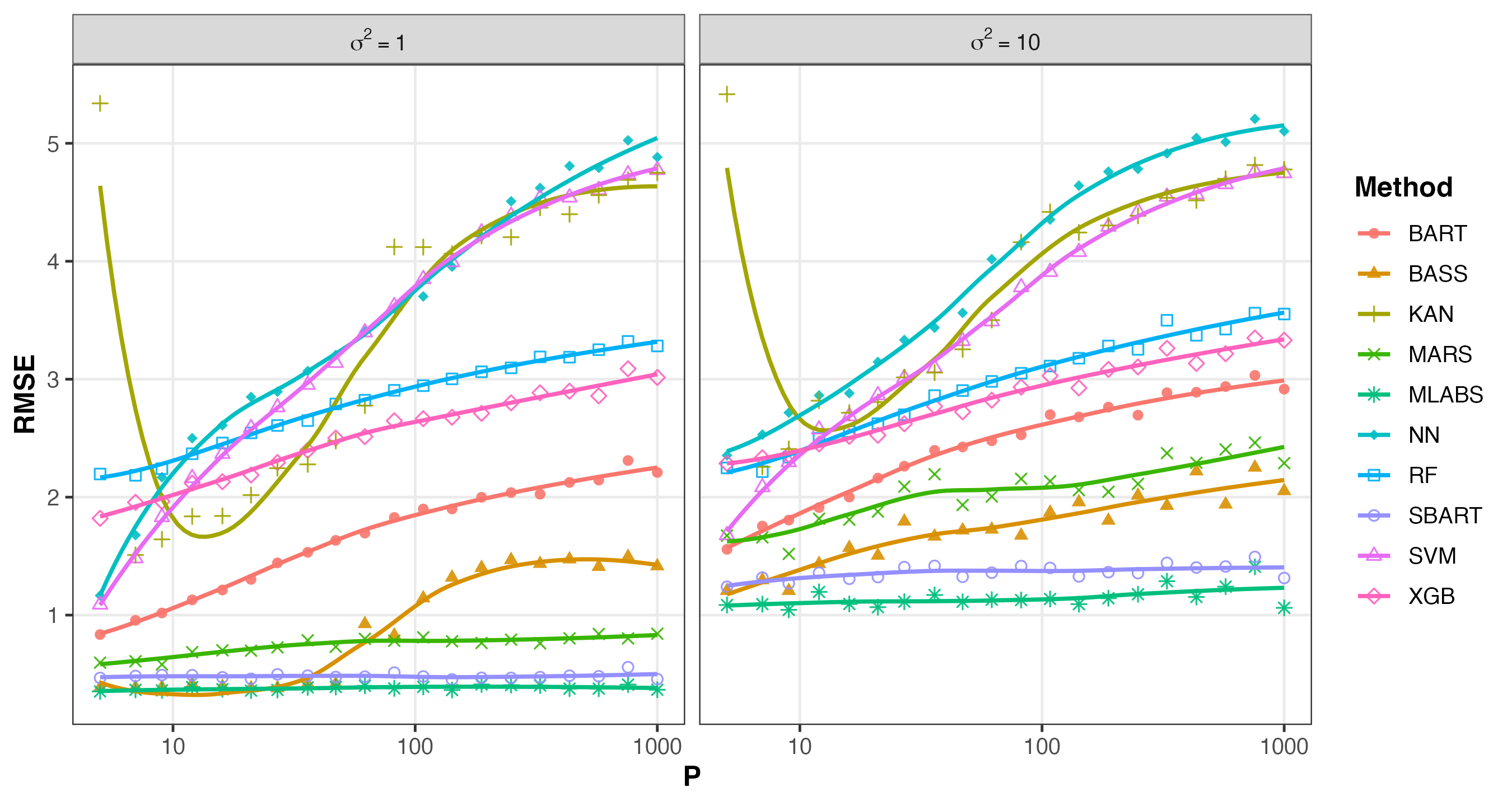}
\caption{
Average test RMSE over 10 replications as a function of the number of
predictors $p$ for the high-dimensional Friedman~1 experiment under the
two noise levels. The horizontal axis is shown on a logarithmic scale.
}
\label{fig:hd_setting}
\end{figure*}

Figure~\ref{fig:hd_setting} summarizes the results. MLABS maintains stable
predictive accuracy as $p$ increases from 5 to 1,000 under both noise
levels and achieves the lowest RMSE throughout the range of dimensions
considered. SBART is also relatively stable, whereas several competing
methods deteriorate as the number of irrelevant predictors increases.
These results indicate that MLABS can preserve predictive accuracy in the
presence of a large number of irrelevant predictors. Additional results on
the selected number of basis functions and computation time are provided
in Appendix~\ref{sec:supp_sim}.

\begin{table*}[ht!]

\centering
\caption{Summary of the six benchmark regression datasets.}
\label{tab:reg_data}
\footnotesize
\begin{tabular}{lccc}
\toprule
Dataset & \# Samples & \# Variables & Source \\
\midrule
Bodyfat
& 252 & 13 & \pkg{mfp} \\
Boston housing
& 506 & 12 & \pkg{MASS} \\
Concrete compressive strength
& 1030 & 8 & UCI \\
Residential building
& 372 & 103 & UCI \\
Tecator meat
& 215 & 100 & \pkg{caret} \\
Chemical manufacturing process
& 152 & 57 & \pkg{AppliedPredictiveModeling} \\
\bottomrule
\end{tabular}

\end{table*}

\section{Real Data Applications}
\label{sec:applications}

We evaluate MLABS on two real-data applications. We first assess predictive
performance on six benchmark regression datasets with different sample
sizes and predictor dimensions. We then consider an environmental-mixture
analysis, where the primary interest is the fitted joint exposure-response
structure.

\subsection{Benchmark Regression Data}
\label{subsec:reg}

We evaluate MLABS on six real regression datasets obtained from the UCI
Machine Learning Repository and several R packages. The datasets vary
substantially in sample size and predictor dimension, as summarized in
Table~\ref{tab:reg_data}. Because MLABS handles quantitative predictors,
categorical predictors and observations with missing values are excluded.
For the Tecator meat and Residential building datasets, we use protein
percentage and actual sales price as the response, respectively. In the
Bodyfat dataset, the apparent recording error in the height of case~42 is
corrected to $69.5$.

We compare MLABS with the eleven competing methods used in the simulation
study. Predictive performance is assessed using 20 repetitions of five-fold
cross-validation. For each repetition, we compute the RMSE on the held-out
folds and report the average RMSE over the 20 repetitions.

Table~\ref{tab:reg_app} reports the results. MLABS achieves the lowest
average rank of 2.00 and ranks among the top three methods on all six
datasets. It attains the lowest RMSE for the Bodyfat and Boston housing
datasets, ranks second for Concrete compressive strength and Residential
building, and ranks third for Tecator meat and Chemical manufacturing
process. These results indicate stable predictive performance across
datasets with substantially different sample sizes and predictor
dimensions.

\begin{sidewaystable*}[p]
\centering
\small
\setlength{\tabcolsep}{4pt}
\renewcommand{\arraystretch}{1.05}

\caption{
Average RMSE of MLABS and the competing methods over 20 repetitions of
five-fold cross-validation. Ranks among the twelve methods are shown in
parentheses, and the lowest RMSE for each dataset is in bold.
}
\label{tab:reg_app}

\begin{adjustbox}{max width=\linewidth, center}
\begin{tabular}{@{}lrrrrrrrrrrrr@{}}
\toprule
Data & MLABS & SBART & BART & BARK & FlexBART & BASS & RF & SVM &
MARS & NN & XGB & KAN \\
\midrule

Bodyfat
& \textbf{4.023 (1)} & 4.085 (2) & 4.186 (6) & 4.171 (5)
& 4.128 (4) & 4.116 (3) & 4.312 (7) & 5.877 (12)
& 4.348 (9) & 5.473 (11) & 4.334 (8) & 4.518 (10) \\

Boston housing
& \textbf{2.911 (1)} & 3.041 (3) & 3.260 (7) & 3.307 (8)
& 3.119 (4) & 3.833 (10) & 3.190 (6) & 3.744 (9)
& 4.168 (11) & 4.184 (12) & 2.983 (2) & 3.159 (5) \\

Concrete compressive strength
& 4.099 (2) & 4.172 (4) & 4.269 (5) & 6.648 (11)
& 4.154 (3) & 5.474 (8) & 4.892 (7) & 5.897 (9)
& 6.243 (10) & 7.305 (12) & \textbf{4.005 (1)} & 4.799 (6) \\

Residential building
& 93.69 (2) & \textbf{84.72 (1)} & 135.55 (8) & 131.55 (5)
& 126.42 (4) & 131.60 (6) & 246.22 (10) & 909.32 (12)
& 121.15 (3) & 341.77 (11) & 208.81 (9) & 132.66 (7) \\

Tecator meat
& 1.2059 (3) & 1.484 (6) & 2.200 (11) & 1.483 (5)
& 2.551 (12) & \textbf{0.965 (1)} & 2.044 (10) & 1.858 (8)
& 0.971 (2) & 1.703 (7) & 1.954 (9) & 1.441 (4) \\

Chemical manufacturing process
& 1.064 (3) & 1.051 (2) & 1.104 (5) & 1.212 (8)
& 1.160 (7) & 1.276 (10) & 1.080 (4) & 1.869 (11)
& 1.256 (9) & 3.779 (12) & \textbf{1.037 (1)} & 1.111 (6) \\

\midrule
Average rank
& \textbf{2.00 (1)} & 3.00 (2) & 7.00 (7.5) & 7.00 (7.5)
& 5.67 (4) & 6.33 (5.5) & 7.33 (9.5) & 10.17 (11)
& 7.33 (9.5) & 10.83 (12) & 5.00 (3) & 6.33 (5.5) \\

\bottomrule
\end{tabular}
\end{adjustbox}

\end{sidewaystable*}

\subsection{Environmental Mixture Analysis}
\label{subsec:mixture}

We next apply MLABS to data from the 2001--2002 National Health and
Nutrition Examination Survey (NHANES). The analysis includes $n=1{,}003$
individuals, with relative leukocyte telomere length (LTL) as the response
and 18 lipid-adjusted serum persistent organic pollutant (POP) measurements
as the exposure mixture
\citep{mitro2016cross,gibson2019overview,jo2026accelerated}. The exposures
comprise 11 polychlorinated biphenyls (PCBs), three dioxins, and four
furans. The POP measurements and LTL are log-transformed and standardized.
Age is standardized, and both standardized age and its squared term are
included as adjustment covariates.

We fit the partially linear model
\begin{equation*}
y_i
=
\bx_i^\top\boldsymbol{\beta}
+
f(\mathbf z_i)
+
\varepsilon_i,
\qquad
\varepsilon_i\sim\mathcal{N}(0,\sigma^2),
\label{eq:model2}
\end{equation*}
where $i=1,\ldots,n$, and $\bx_i$ contains an intercept, standardized age, and squared
standardized age, and $\mathbf z_i=(z_{i1},\ldots,z_{i18})^\top$ denotes the POP mixture. The age terms enter linearly, while the joint exposure-response function
$f$ is modeled nonparametrically.

We compare MLABS with Bayesian kernel machine regression
(BKMR; \citealp{bobb2015bayesian}), a widely used method for environmental
mixture analysis. Table~\ref{tab:cvm} reports five-fold cross-validated
predictive accuracy and computation time. MLABS and BKMR achieve comparable
predictive accuracy, while MLABS requires substantially less computation
time.

\begin{table}[!ht]
\caption{
Five-fold cross-validated predictive performance for the NHANES POP--LTL
data. RMSE is reported as the mean (standard deviation) across folds, and
computing time is averaged across folds.
}
\label{tab:cvm}
\centering
\begin{tabular}{lcc}
\toprule
Method & RMSE & Computing time (sec.) \\
\midrule
BKMR  & 0.8747 (0.0392) & 3249 \\
MLABS & 0.8781 (0.0388) & 700.6 \\
\bottomrule
\end{tabular}

\end{table}

To characterize the fitted mixture-response surface, let
$\mathbf z^{(q)}=(z_1^{(q)},\ldots,z_{18}^{(q)})^\top$, where
$z_j^{(q)}$ denotes the empirical $q$th quantile of exposure $j$. We define
the overall mixture contrast relative to the median exposure profile as
\[
\Delta(q)
=
f\{\mathbf z^{(q)}\}
-
f\{\mathbf z^{(0.5)}\}.
\]
Thus, $\Delta(q)$ summarizes the change in the fitted LTL response when all
POP exposures are jointly shifted from their respective medians to their
$q$th quantiles. Because the POPs are correlated, these joint-quantile profiles may represent
exposure combinations that are uncommon in the observed data. We therefore
interpret $\Delta(q)$ as a descriptive summary of the fitted response
surface rather than as a causal effect.

\begin{figure}[t]
\centering

\begin{minipage}[t]{0.48\textwidth}
\centering
\includegraphics[width=\textwidth]
{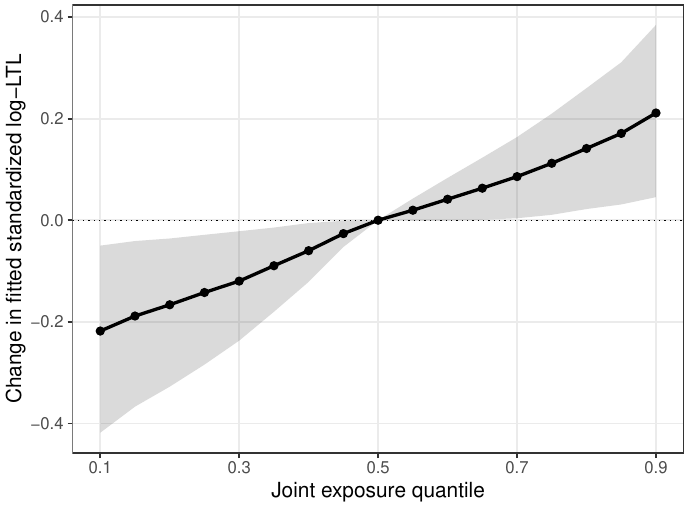}

\smallskip
{\small (a) Overall mixture association.}
\end{minipage}
\hfill
\begin{minipage}[t]{0.48\textwidth}
\centering
\includegraphics[width=\textwidth]
{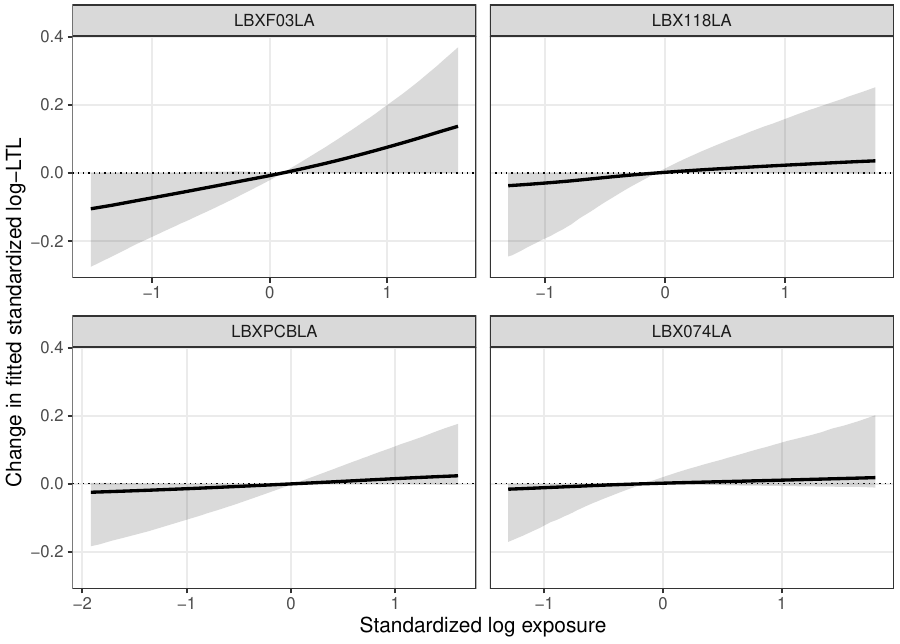}

\smallskip
{\small (b) Selected univariate exposure-response functions.}
\end{minipage}

\caption{
Posterior summaries of the fitted exposure-response function for the NHANES
POP--LTL data. Panel (a) shows the overall mixture contrast $\Delta(q)$, and
Panel (b) shows the four exposure-response functions with the largest
posterior-mean variation. Solid lines denote posterior means and shaded
regions denote pointwise 95\% credible intervals.
}
\label{fig:mixture_effects}
\end{figure}

Figure~\ref{fig:mixture_effects} summarizes the posterior mixture-response
patterns. The posterior mean of $\Delta(q)$ increases across the exposure
distribution, from approximately $-0.18$ at the 10th marginal quantiles to
$0.16$ at the 90th marginal quantiles. Although the pointwise 95\% credible
intervals are relatively wide, the posterior probability that the fitted
response increases from the 10th to the 90th joint exposure quantile is
approximately $0.95$. Thus, the results suggest a positive overall association between the joint POP exposure profile and LTL, although the magnitude of this association remains uncertain.

We also examine exposure-specific response functions by varying one exposure
at a time and holding the remaining exposures at their medians. The
individual exposure-response functions exhibit substantial uncertainty.
Among the 18 exposures, LBXF03LA shows the most pronounced posterior-mean
response pattern, whereas the remaining exposures show smaller changes.
The pointwise credible bands are nevertheless relatively wide and include
zero over much of the exposure range. Overall, the increasing pattern is
more evident for the joint mixture than for any individual POP.

\section{Discussion}
\label{sec:discussion}

We proposed Multivariate L\'evy Adaptive B-Spline Regression (MLABS), a
Bayesian nonparametric regression model based on sparse random expansions
of tensor-product B-spline basis functions. MLABS jointly learns the basis
structure, including active predictors, interaction orders, spline degrees,
and knot locations. This flexible representation can accommodate smooth,
localized, and discontinuous features without requiring a fixed
tensor-product basis expansion.

The numerical studies demonstrate competitive predictive performance
across a range of regression settings. MLABS performs consistently well for
the simulated functions with different local structures and maintains
predictive accuracy in the presence of many irrelevant predictors. It also
ranks among the top-performing methods across the benchmark real datasets.
In the environmental-mixture application, MLABS achieves predictive
accuracy comparable to BKMR with substantially less computation and
provides posterior summaries of both joint and exposure-specific response
patterns.

Several limitations remain. The reversible-jump MCMC algorithm can become
computationally demanding as the complexity of the basis representation
increases, and the current formulation is restricted to quantitative
predictors. One possible computational direction is to replace
trans-dimensional exploration with a sufficiently rich dictionary of
tensor-product B-spline basis functions and sparsity-inducing priors.
Extensions to mixed-type predictors are also of practical interest. On the
theoretical side, recent work has established posterior contraction rates
for univariate LABS over Besov classes \citep{oh2026posterior}. An important
direction is to extend these results to MLABS over multivariate Besov
classes with low-order interaction structure.

\bibliographystyle{dcu}
\bibliography{sn-bibliography}

@phdthesis{tu2006bayesian,
  title={Bayesian nonparametric modeling using {L}{\`e}vy process priors with applications for function estimation, time series modeling and spatio-temporal modeling},
  author={Tu, Chong},
  year={2006},
  type     = {{PhD} dissertation},
  school={Duke University}
}

@article{park2023levy,
  title={L{\'e}vy adaptive {B}-spline regression via overcomplete systems},
  author={Park, Sewon and Oh, Hee-Seok and Lee, Jaeyong},
  journal={Statistica Sinica},
  volume={33},
  number={4},
  pages={2715--2737},
  year={2023},
  publisher={JSTOR}
}

@article{green1995reversible,
  title={Reversible jump Markov chain Monte Carlo computation and Bayesian model determination},
  author={Green, Peter J},
  journal={Biometrika},
  volume={82},
  number={4},
  pages={711--732},
  year={1995},
  publisher={Oxford University Press},
  doi={10.2307/2337340}
}

@article{nott2005efficient,
  title={Efficient sampling schemes for Bayesian {MARS} models with many predictors},
  author={Nott, David J and Kuk, Anthony YC and Duc, Hiep},
  journal={Statistics and Computing},
  volume={15},
  number={2},
  pages={93--101},
  year={2005},
  publisher={Springer},
  doi={10.1007/s11222-005-6201-x}
}

@inproceedings{imaizumi2019deep,
  title={Deep neural networks learn non-smooth functions effectively},
  author={Imaizumi, Masaaki and Fukumizu, Kenji},
  booktitle={The 22nd International Conference on Artificial Intelligence and Statistics},
  pages={869--878},
  year={2019},
  organization={PMLR}
}

@article{francom2018sensitivity,
  title={Sensitivity analysis and emulation for functional data using Bayesian adaptive splines},
  author={Francom, Devin and Sans{\'o}, Bruno and Kupresanin, Ana and Johannesson, Gardar},
  journal={Statistica Sinica},
  volume={28},
  pages={791--816},
  year={2018}
}

@article{hwang1994regression,
  title={Regression modeling in back-propagation and projection pursuit learning},
  author={Hwang, Jeng-Neng and Lay, Shyh-Rong and Maechler, Martin and Martin, R Douglas and Schimert, Jim},
  journal={IEEE Transactions on neural networks},
  volume={5},
  number={3},
  pages={342--353},
  year={1994},
  publisher={IEEE},
  doi={10.1109/72.286906}
}

@article{friedman1991multivariate,
  title={Multivariate adaptive regression splines},
  author={Friedman, Jerome H},
  journal={The Annals of Statistics},
  pages={1--67},
  year={1991},
  publisher={JSTOR}
}

@phdthesis{ouyang2008bayesian,
  title={Bayesian additive regression kernels},
  author={Ouyang, Zhi},
  year={2008},
  type     = {{PhD} dissertation},
  school={Duke University}
}

@inproceedings{chen2016xgboost,
  title={{XGB}oost: A scalable tree boosting system},
  author={Chen, Tianqi and Guestrin, Carlos},
  booktitle={Proceedings of the 22nd acm sigkdd international conference on knowledge discovery and data mining},
  pages={785--794},
  year={2016},
  doi={10.1145/2939672.2939785}
}

@article{chipman2010bart,
  title={{BART}: Bayesian additive regression trees},
  author={Chipman, Hugh A and George, Edward I and McCulloch, Robert E},
  journal={The Annals of Applied Statistics},
  volume={4},
  number={1},
  pages={266--298},
  year={2010},
  publisher={Institute of Mathematical Statistics},
  doi={10.1214/09-AOAS285}
}

@article{breiman2001random,
  title={Random forests},
  author={Breiman, Leo},
  journal={Machine learning},
  volume={45},
  number={1},
  pages={5--32},
  year={2001},
  publisher={Springer},
  doi = {10.1023/A:1010933404324}
}

@article{linero2018sbart,
  title={Bayesian regression tree ensembles that adapt to smoothness and sparsity},
  author={Linero, Antonio R and Yang, Yun},
  journal={Journal of the Royal Statistical Society Series B: Statistical Methodology},
  volume={80},
  number={5},
  pages={1087--1110},
  year={2018},
  publisher={Wiley Online Library},
  doi={10.1111/rssb.12293}
}

@article{linero2018dart,
  title={Bayesian regression trees for high-dimensional prediction and variable selection},
  author={Linero, Antonio R},
  journal={Journal of the American Statistical Association},
  volume={113},
  number={522},
  pages={626--636},
  year={2018},
  publisher={Taylor \& Francis},
  doi={10.1080/01621459.2016.1264957}
}

@article{meyer2015support,
  title={Support vector machines},
  author={Meyer, David and Wien, FH Technikum},
  journal={The Interface to libsvm in package e1071},
  volume={28},
  year={2015}
}

@article{francom2020bass,
  title={BASS: An R package for fitting and performing sensitivity analysis of bayesian adaptive spline surfaces},
  author={Francom, Devin and Sans{\'o}, Bruno},
  journal={Journal of Statistical Software},
  volume={94},
  number={1},
  pages={1--36},
  year={2020}
}

@article{hastie2000bayesian,
  title={Bayesian backfitting (with comments and a rejoinder by the authors)},
  author={Hastie, Trevor and Tibshirani, Robert and others},
  journal={Statistical Science},
  volume={15},
  number={3},
  pages={196--223},
  year={2000},
  publisher={Institute of Mathematical Statistics},
  doi={10.1214/ss/1009212815}
}

@article{kimeldorf1971some,
  title={Some results on Tchebycheffian spline functions},
  author={Kimeldorf, George and Wahba, Grace},
  journal={Journal of mathematical analysis and applications},
  volume={33},
  number={1},
  pages={82--95},
  year={1971},
  publisher={Academic Press},
  doi={10.1016/0022-247X(71)90184-3}
}

@book{wahba1990spline,
  title={Spline models for observational data},
  author={Wahba, Grace},
  volume={59},
  year={1990},
  publisher={{SIAM}},
  address = {Philadelphia, PA}
}

@article{denison1998bayesian,
  title={Bayesian {MARS}},
  author={Denison, David GT and Mallick, Bani K and Smith, Adrian FM},
  journal={Statistics and Computing},
  volume={8},
  number={4},
  pages={337--346},
  year={1998},
  publisher={Springer},
  doi={10.1023/A:1008824606259}
}

@article{friedman2001greedy,
  title={Greedy function approximation: a gradient boosting machine},
  author={Friedman, Jerome H},
  journal={Annals of Statistics},
  pages={1189--1232},
  year={2001},
  publisher={JSTOR},
  doi={10.1214/aos/1013203451}
}

@article{liu2024kan,
  title={{KAN}: Kolmogorov--{A}rnold Networks},
  author={Liu, Ziming and Wang, Yixuan and Vaidya, Sachin and Ruehle, Fabian and 
          Halverson, James and Solja{\v{c}}i{\'c}, Marin and Hou, Thomas Y and 
          Tegmark, Max},
  journal={arXiv preprint arXiv:2404.19756},
  year={2024},
  doi={10.48550/arXiv.2404.19756}
}

@article{linero2025generalized,
  title={Generalized Bayesian additive regression trees models: Beyond conditional conjugacy},
  author={Linero, Antonio R},
  journal={Journal of the American Statistical Association},
  volume={120},
  number={549},
  pages={356--369},
  year={2025},
  publisher={Taylor \& Francis},
  doi={10.1080/01621459.2024.2337156}
}

@article{bobb2015bayesian,
  title={Bayesian kernel machine regression for estimating the health effects of multi-pollutant mixtures},
  author={Bobb, Jennifer F and Valeri, Linda and Claus Henn, Birgit and Christiani, David C and Wright, Robert O and Mazumdar, Maitreyi and Godleski, John J and Coull, Brent A},
  journal={Biostatistics},
  volume={16},
  number={3},
  pages={493--508},
  year={2015},
  publisher={Oxford University Press},
  doi={10.1093/biostatistics/kxu058}
}

@article{dimatteo2001bayesian,
  title={Bayesian curve-fitting with free-knot splines},
  author={DiMatteo, Ilaria and Genovese, Christopher R and Kass, Robert E},
  journal={Biometrika},
  volume={88},
  number={4},
  pages={1055--1071},
  year={2001},
  publisher={Oxford University Press},
  doi={10.1093/biomet/88.4.1055}
}

@article{linero2022softbart,
  title={Softbart: Soft Bayesian additive regression trees},
  author={Linero, Antonio R},
  journal={arXiv preprint arXiv:2210.16375},
  year={2022}
}

@book{schumaker2007spline,
  title={Spline functions: basic theory},
  author={Schumaker, Larry},
  year={2007},
  publisher={Cambridge university press}
}

@inproceedings{genz1984testing,
  title={Testing multidimensional integration routines},
  author={Genz, Alan},
  booktitle={Proc. of international conference on Tools, methods and languages for scientific and engineering computation},
  pages={81--94},
  year={1984}
}

@article{mitro2016cross,
  title={Cross-sectional associations between exposure to persistent organic pollutants and leukocyte telomere length among US adults in {NHANES}, 2001--2002},
  author={Mitro, Susanna D and Birnbaum, Linda S and Needham, Belinda L and Zota, Ami R},
  journal={Environmental health perspectives},
  volume={124},
  number={5},
  pages={651--658},
  year={2016},
  publisher={ACS Publications},
  doi={10.1289/ehp.1510187}
}

@article{gibson2019overview,
  title={An overview of methods to address distinct research questions on environmental mixtures: an application to persistent organic pollutants and leukocyte telomere length},
  author={Gibson, Elizabeth A and Nunez, Yanelli and Abuawad, Ahlam and Zota, Ami R and Renzetti, Stefano and Devick, Katrina L and Gennings, Chris and Goldsmith, Jeff and Coull, Brent A and Kioumourtzoglou, Marianthi-Anna},
  journal={Environmental Health},
  volume={18},
  number={1},
  pages={76},
  year={2019},
  publisher={Springer},
  doi={10.1186/s12940-019-0515-1}
}

@article{jo2026accelerated,
  title={Accelerated Bayesian Kernel Machine Regression: A Gaussian Variational Approximation with the Horseshoe Prior},
  author={Jo, Seongil and Ye, Shinhee and Hahn, Georg and Lee, Woojoo},
  journal={Statistics and Computing},
  volume={36},
  number={2},
  pages={79},
  year={2026},
  publisher={Springer},
  doi={10.1007/s11222-026-10831-x}
}

@article{wolpert2011stochastic,
  title={Stochastic expansions using continuous dictionaries: L{\'e}vy adaptive regression kernels},
  author={Wolpert, Robert L and Clyde, Merlise A and Tu, Chong},
  journal={The Annals of Statistics},
  pages={1916--1962},
  year={2011},
  publisher={JSTOR},
  doi={10.1214/11-AOS889}
}

@article{oh2026posterior,
  title={Posterior Contraction of {L}{\'e}vy Adaptive {B}-spline Regression in {B}esov Spaces},
  author={Oh, Jeunghun and Park, Sewon and Lee, Jaeyong},
  journal={arXiv preprint arXiv:2605.19610},
  year={2026},
  doi={10.48550/arXiv.2605.19610}
}

@article{sande2019sharp,
  title={Sharp error estimates for spline approximation: Explicit constants, n-widths, and eigenfunction convergence},
  author={Sande, Espen and Manni, Carla and Speleers, Hendrik},
  journal={Mathematical Models and Methods in Applied Sciences},
  volume={29},
  number={06},
  pages={1175--1205},
  year={2019},
  publisher={World Scientific}
}

@book{triebel1983theory,
  title={Theory of Function Spaces},
  author={Triebel, Hans},
  volume={78},
  series={Monographs in Mathematics},
  year={1983},
  publisher={Birkh{\"a}user},
  address={Basel, Boston, Stuttgart},
  doi={10.1007/978-3-0346-0416-1},
  isbn={978-3-7643-1381-4}
}

\newpage
\begin{appendices}

\section{Computational Details}
\label{sec:supp_comp}

This section provides additional details on posterior computation for
MLABS. We describe the within-model backfitting updates, the joint
degree--knot refresh moves, the residual-guided structural proposals,
the reversible-jump acceptance ratios, and the global parameter updates.

\subsection{Notation and target posterior}
\label{subsec:supp_notation}

Write
\[
\Theta_J
=
\{(\beta_j,\boldsymbol{\psi}_j)\}_{j=1}^J,
\qquad
\boldsymbol{\psi}_j
=
(K_j,\boldsymbol{\nu}^{(j)},
 \boldsymbol{c}^{(j)},\boldsymbol{\xi}^{(j)}),
\]
where $\beta_j$ is the coefficient of the $j$th tensor-product B-spline
basis, $K_j$ is its interaction order,
$\boldsymbol{\nu}^{(j)}$ is the vector of active predictors,
$\boldsymbol{c}^{(j)}$ is the vector of B-spline degrees, and
$\boldsymbol{\xi}^{(j)}$ denotes the corresponding knot state.

Let
\[
\boldsymbol b_j
=
\left(
\mathbf B_j(\mathbf x_1;\boldsymbol{\psi}_j),
\ldots,
\mathbf B_j(\mathbf x_n;\boldsymbol{\psi}_j)
\right)^\top,
\]
and define
\[
\mathbf B
=
[\boldsymbol b_1,\ldots,\boldsymbol b_J].
\]
The fitted mean vector at the current MCMC state is
\[
\boldsymbol\mu_J
=
\beta_0\mathbf1_n+\mathbf B\boldsymbol\beta.
\]
The intercept $\beta_0$ is fixed at the sample mean of the response
throughout posterior sampling. When the response is standardized,
this corresponds to $\beta_0=0$.

We condition on the observed predictor design
$\mathbf X=(\mathbf x_1,\ldots,\mathbf x_n)^\top$ throughout.  The
odd-degree knot prior introduced below uses the empirical distribution of
the corresponding predictor, but it does not depend on the response or on
the residuals.  Conditional on $\mathbf X$ and $\beta_0$, the target
posterior is
\begin{equation}
\label{eq:supp_post_kernel}
\begin{aligned}
&\pi(
\Theta_J,J,M,\sigma^2,\boldsymbol\pi
\mid\boldsymbol Y,\mathbf X,\beta_0)
\\
&\qquad\propto
L(\boldsymbol Y\mid\Theta_J,J,\sigma^2,\beta_0)
\,
\pi(\Theta_J\mid J,M,\boldsymbol\pi,\mathbf X)
\,
\pi(J\mid M,\lambda_J)
\,
\pi(M)
\,
\pi(\sigma^2)
\,
\pi(\boldsymbol\pi).
\end{aligned}
\end{equation}

Let
\[
\bar J_M=Me^{-\lambda_J},
\qquad
\tau
=
c_{\mathrm{shrink}}\operatorname{sd}(\boldsymbol Y).
\]
The coefficient prior is
\begin{equation}
\label{eq:supp_phiM}
\beta_j\mid M
\stackrel{\mathrm{iid}}{\sim}
N(0,\phi_M^2),
\qquad
\phi_M^2
=
\frac{\tau^2}{\bar J_M}
=
\frac{\tau^2}{Me^{-\lambda_J}}.
\end{equation}
Thus, conditional on $M$, the coefficient scale depends on the
expected number of basis functions rather than on the realized number
$J$. Consequently, the prior scale of the coefficients associated
with existing basis functions remains unchanged under birth and death
moves.

\subsection{Within-model backfitting updates}
\label{subsec:supp_within}

For each basis function, the coefficient and local knot state are
updated conditionally on the remaining basis terms.

\paragraph{Coefficient update.}
For basis function $j$, define the partial residual
\[
\boldsymbol r^{(j)}
=
\boldsymbol Y
-
\beta_0\mathbf1_n
-
\sum_{k\ne j}
\boldsymbol b_k\beta_k.
\]
Conditional on the remaining parameters,
\[
\boldsymbol r^{(j)}
=
\boldsymbol b_j\beta_j+\boldsymbol\varepsilon,
\qquad
\boldsymbol\varepsilon
\sim
N(\mathbf0,\sigma^2\mathbf I_n).
\]
Combining the Gaussian likelihood with
$\beta_j\sim N(0,\phi_M^2)$ gives
\begin{equation}
\label{eq:supp_beta_fc}
\beta_j\mid\mathrm{rest}
\sim
N(m_j,v_j^2),
\end{equation}
where
\begin{equation}
\label{eq:supp_vj}
v_j^2
=
\left(
\frac{\boldsymbol b_j^\top\boldsymbol b_j}{\sigma^2}
+
\frac{1}{\phi_M^2}
\right)^{-1},
\qquad
m_j
=
v_j^2
\frac{\boldsymbol b_j^\top\boldsymbol r^{(j)}}{\sigma^2}.
\end{equation}

\paragraph{Local continuous knot update.}
Conditional on $\beta_j$, knot coordinates are updated locally while the
interaction order, active predictors, and spline degrees remain fixed.
For an even-degree factor, all knot coordinates are updated.  For an
odd-degree factor, the continuous central knot is held fixed in this local
move and only the side knots are updated; the central knot and both sets of
side knots are updated jointly in the separate move described below.

For a factor associated with predictor $\ell$, independent Gaussian
increments with standard deviation
$s_\xi(b_\ell-a_\ell)$ are added to the coordinates included in the local
move.  The resulting knot vector is sorted.  A proposal is rejected
immediately unless it lies strictly inside $(a_\ell,b_\ell)$ and is
strictly ordered.  For an odd-degree factor, it is also rejected unless
the fixed central value remains in the central position after sorting.
These checks are the ones used in the implementation.

On the feasible ordered region the random-walk proposal is symmetric.  For
an even degree, the knot prior is uniform on that region.  For an odd
degree, conditional on the fixed central knot $A$, the density of the side
knots is constant on the corresponding left and right ordered regions.
Thus, the prior and proposal terms cancel in either case.

Let $\boldsymbol b_j'$ denote the basis vector corresponding to the
proposed knot state.  The proposal is accepted with probability
$\min\{1,A_{\mathrm{loc}}\}$, where
\begin{equation}
\label{eq:supp_local_knot}
\log A_{\mathrm{loc}}
=
-\frac{1}{2\sigma^2}
\left[
\left\|
\boldsymbol r^{(j)}-\beta_j\boldsymbol b_j'
\right\|^2
-
\left\|
\boldsymbol r^{(j)}-\beta_j\boldsymbol b_j
\right\|^2
\right].
\end{equation}

\paragraph{Central-knot update for odd degrees.}
Consider an odd-degree factor of degree $c$ associated with predictor
$\ell$.  Write
\[
k_c=\frac{c+1}{2},
\qquad
m_c=\frac{c+3}{2},
\qquad
A=\xi_{m_c}.
\]
The central knot $A$ is continuous.  Its prior density is the
design-based density $h_\ell$ defined in
\eqref{eq:supp_odd_prior_center}, and its residual-guided proposal density
is $g_\ell(\cdot\mid\boldsymbol r^{(j)})$ defined in
\eqref{eq:supp_odd_proposal_center}.  The partial residual
$\boldsymbol r^{(j)}$ excludes basis function $j$, so it remains unchanged
when this factor is modified.

A proposed center is drawn as
\[
A'\sim g_\ell(\cdot\mid\boldsymbol r^{(j)}).
\]
Conditional on $A'$, draw $k_c$ values from $U(a_\ell,A')$ and $k_c$
values from $U(A',b_\ell)$ and sort each group.  These draws form the new
side knots.  Let
\begin{align}
\begin{split}
\label{eq:supp_side_density}
p_{\mathrm{side}}(\boldsymbol\xi_{-m_c}\mid A)  =
\frac{(k_c!)^2}
{(A-a_\ell)^{k_c}(b_\ell-A)^{k_c}}
\mathbf 1\!\left\{
a_\ell<\xi_1<\cdots<\xi_{m_c-1}<A
<\xi_{m_c+1}<\cdots<\xi_{c+2}<b_\ell
\right\}.
\end{split}
\end{align}
The forward and reverse proposal densities are, respectively,
\[
\begin{aligned}
q_{\mathrm{fwd}}
&=
g_\ell(A'\mid\boldsymbol r^{(j)})
p_{\mathrm{side}}(\boldsymbol\xi_{-m_c}'\mid A'),
\\
q_{\mathrm{rev}}
&=
g_\ell(A\mid\boldsymbol r^{(j)})
p_{\mathrm{side}}(\boldsymbol\xi_{-m_c}\mid A).
\end{aligned}
\]
The corresponding knot-prior densities are obtained by replacing $g_\ell$
with $h_\ell$.  Hence the side-knot densities cancel exactly.

Let $\boldsymbol b_j'$ denote the basis vector after replacing this factor
by its proposed knot vector.  Conditional on the current coefficient
$\beta_j$, the move is accepted with probability
$\min\{1,A_{\mathrm{cen}}\}$, where
\begin{equation}
\label{eq:supp_anchor_update}
\begin{aligned}
\log A_{\mathrm{cen}}
=&
-\frac{1}{2\sigma^2}
\left[
\left\|
\boldsymbol r^{(j)}-\beta_j\boldsymbol b_j'
\right\|^2
-
\left\|
\boldsymbol r^{(j)}-\beta_j\boldsymbol b_j
\right\|^2
\right]
+
\log\frac{h_\ell(A')}{h_\ell(A)}
+
\log\frac{
g_\ell(A\mid\boldsymbol r^{(j)})
}{
g_\ell(A'\mid\boldsymbol r^{(j)})
}.
\end{aligned}
\end{equation}
All quantities in this update are densities with respect to Lebesgue
measure; no discrete anchor index is part of the Markov state.

\subsection{Collapsed marginal score and joint degree--knot refresh}
\label{subsec:supp_refresh}

Structural changes are evaluated with the corresponding basis
coefficient integrated out. Let $\boldsymbol r$ denote a residual
vector to be explained by a candidate basis vector $\boldsymbol b$.
Under
\[
\boldsymbol r\mid\beta
\sim
N(\beta\boldsymbol b,\sigma^2\mathbf I_n),
\qquad
\beta\sim N(0,\phi_M^2),
\]
define
\begin{equation}
\label{eq:supp_collapsed_moments}
v^2
=
\left(
\frac{\boldsymbol b^\top\boldsymbol b}{\sigma^2}
+
\frac{1}{\phi_M^2}
\right)^{-1},
\qquad
m
=
v^2
\frac{\boldsymbol b^\top\boldsymbol r}{\sigma^2}.
\end{equation}
Integrating out $\beta$ gives the log marginal contribution of the
candidate basis, relative to omitting it,
\begin{equation}
\label{eq:supp_collapsed_score}
\ell(\boldsymbol b;\boldsymbol r)
=
\frac12
\log\left(\frac{v^2}{\phi_M^2}\right)
+
\frac12\frac{m^2}{v^2},
\end{equation}
up to an additive constant independent of $\boldsymbol b$.

Let $n_{\mathrm{ref}}$ be a fixed tuning parameter specifying the
maximum number of basis functions subjected to a refresh move in each
MCMC iteration. The default implementation uses
$n_{\mathrm{ref}}=5$. After the within-model backfitting sweep, the
actual number of basis functions selected for refresh is
\[
n_{\mathrm{ref}}^\ast
=
\min(n_{\mathrm{ref}},J).
\]
A set of $n_{\mathrm{ref}}^\ast$ distinct basis functions is then
selected uniformly without replacement. For each selected basis $j$,
the interaction order $K_j$ and active predictors
$\boldsymbol{\nu}^{(j)}$ are held fixed, while
\[
\boldsymbol{c}^{(j)}
\rightarrow
\boldsymbol{c}^{(j)\prime},
\qquad
\boldsymbol{\xi}^{(j)}
\rightarrow
\boldsymbol{\xi}^{(j)\prime}
\]
are proposed jointly. Joint updating ensures that each proposed
degree-$c$ factor is accompanied by a knot state of the appropriate
dimension.

Each proposed degree is drawn uniformly from the allowed degree set
$S$, which coincides with the degree prior. Conditional on the
proposed degree, a new knot state is generated from the
residual-guided knot proposal of
Section~\ref{subsec:supp_knot}, using the partial residual
$\boldsymbol r^{(j)}$.

Let
\[
q_{\mathrm{knot}}
(
\boldsymbol\xi
\mid
\boldsymbol r,
\boldsymbol\nu,
\boldsymbol c)
\]
denote the product of the degree-specific knot proposal densities over
the active factors. With $\boldsymbol b_j$ and $\boldsymbol b_j'$
denoting the current and proposed basis vectors, the refresh move is
accepted with probability $\min\{1,A_{\mathrm{ref}}\}$, where
\begin{equation}
\label{eq:supp_refresh}
\begin{aligned}
\log A_{\mathrm{ref}}
=&\;
\ell(\boldsymbol b_j';\boldsymbol r^{(j)})
-
\ell(\boldsymbol b_j;\boldsymbol r^{(j)})
\\
&+
\log
\frac{
\pi_\xi(
\boldsymbol\xi^{(j)\prime}
\mid
\boldsymbol c^{(j)\prime},
\boldsymbol\nu^{(j)})
}{
\pi_\xi(
\boldsymbol\xi^{(j)}
\mid
\boldsymbol c^{(j)},
\boldsymbol\nu^{(j)})
}
\\
&+
\log
\frac{
q_{\mathrm{knot}}
(
\boldsymbol\xi^{(j)}
\mid
\boldsymbol r^{(j)},
\boldsymbol\nu^{(j)},
\boldsymbol c^{(j)})
}{
q_{\mathrm{knot}}
(
\boldsymbol\xi^{(j)\prime}
\mid
\boldsymbol r^{(j)},
\boldsymbol\nu^{(j)},
\boldsymbol c^{(j)\prime})
}.
\end{aligned}
\end{equation}

For an even degree $c$, define
\[
E_{c,\ell}(\boldsymbol\xi;\boldsymbol r)
=
\frac{
\dfrac{(c+2)!}{(b_\ell-a_\ell)^{c+2}}
}{
\displaystyle
\sum_{
s:\,
\xi_{k_c}<x_{s\ell}<\xi_{k_c+1}
}
w_s(\boldsymbol r;\ell)
\frac{(k_c!)^2}
{
(x_{s\ell}-a_\ell)^{k_c}
(b_\ell-x_{s\ell})^{k_c}
}
},
\qquad
k_c=\frac{c+2}{2}.
\]

For an odd degree $c$, define
\begin{equation}
\label{eq:supp_E_odd}
E_{c,\ell}(\boldsymbol\xi;\boldsymbol r)
=
\frac{
h_\ell(\xi_{m_c})
}{
g_\ell(\xi_{m_c}\mid\boldsymbol r)
},
\qquad
m_c=\frac{c+3}{2}.
\end{equation}
The conditional side-knot density cancels from this ratio.  Therefore,
the knot part of the refresh ratio in 
\eqref{eq:supp_refresh} can equivalently be evaluated as the product of
$E_{c,\ell}$ values for the proposed factors divided by the corresponding
product for the current factors, using the even- or odd-degree definition
as appropriate.

The degree-prior and degree-proposal factors cancel because both are
uniform on $S$.  For every allowed degree the complete knot vector has a
density with respect to Lebesgue measure on its ordered-knot space.  No
Jacobian term is required because a new knot state is sampled directly on
the proposed degree-specific space rather than being obtained through a
deterministic dimension-matching transformation.

If the refresh move is accepted, $\beta_j$ is resampled from
$N(m,v^2)$ using \eqref{eq:supp_collapsed_moments} with the proposed
basis vector.

\subsection{Residual relevance score and variable proposal}
\label{subsec:supp_relevance}

Given a residual vector
\[
\boldsymbol r=(r_1,\ldots,r_n)^\top,
\qquad
\bar r=n^{-1}\sum_{i=1}^n r_i,
\]
consider predictor $\ell\in\{1,\ldots,p\}$. After scaling the
predictor to $[0,1]$, partition its range into
\[
B
=
\min\!\left\{n,20,\max\{5,\lfloor\sqrt n\rfloor\}\right\}
\]
equal-width bins, with empty bins ignored. Let
$\mathcal I_{\ell b}$ denote the observations in bin $b$, and define
\[
\bar r_{\ell b}
=
\frac{1}{|\mathcal I_{\ell b}|}
\sum_{i\in\mathcal I_{\ell b}}r_i.
\]
Let
\[
SS_{\mathrm{total}}
=
\sum_{i=1}^n(r_i-\bar r)^2
\]
and
\[
SS_{\mathrm{within},\ell}
=
\sum_{b:\,|\mathcal I_{\ell b}|>0}
\sum_{i\in\mathcal I_{\ell b}}
(r_i-\bar r_{\ell b})^2.
\]
When $SS_{\mathrm{total}}>0$, define
\begin{equation}
\label{eq:supp_relevance}
\rho_\ell
=
\max\left\{
0,\,
1-
\frac{
SS_{\mathrm{within},\ell}
}{
SS_{\mathrm{total}}
}
\right\},
\end{equation}
and set $\rho_\ell=0$ when $SS_{\mathrm{total}}=0$.

Let
$\boldsymbol\pi=(\pi_1,\ldots,\pi_p)$ denote the current
variable-selection probabilities. We form
\begin{equation}
\label{eq:supp_informed}
\widetilde\pi_\ell
=
\frac{
\pi_\ell(1+\kappa\rho_\ell)
}{
\sum_{h=1}^p
\pi_h(1+\kappa\rho_h)
},
\qquad
\ell=1,\ldots,p,
\end{equation}
where $\kappa\ge0$ controls the strength of the residual guidance.
Conditional on interaction order $K$, the active predictors are then
sampled without replacement according to
\[
q_\nu(
\boldsymbol\nu\mid K,\widetilde{\boldsymbol\pi})
=
p_{\mathrm{WOR}}
(
\boldsymbol\nu\mid K,\widetilde{\boldsymbol\pi}),
\]
where $p_{\mathrm{WOR}}$ is the weighted
sampling-without-replacement probability defined in the main paper.

For a birth move, the relevance scores are computed from the residual
under the current $J$-basis model. For a death move, the reverse birth
proposal is evaluated using the partial residual obtained after
removing the basis selected for deletion.

The residual-guided probabilities
$\widetilde{\boldsymbol{\pi}}$ are used only in the variable proposal
distribution. The variable prior remains
$p_{\mathrm{WOR}}(\boldsymbol{\nu}\mid K,\boldsymbol{\pi})$, and the
difference between the prior and proposal probabilities is accounted
for in the Metropolis--Hastings acceptance ratio. Thus, residual
guidance affects the efficiency of posterior exploration but does not
change the target posterior distribution.

\subsection{Residual-guided knot proposal}
\label{subsec:supp_knot}

Fix an active predictor $\ell$ and a residual vector
$\boldsymbol r$.  Its knot domain $(a_\ell,b_\ell)$ is a slightly
expanded interval containing the observed predictor range.  Let
\[
B_\xi
=
\min\!\left\{n,20,
\max\{5,\lfloor\sqrt n\rfloor\}\right\},
\]
and partition the observed range of predictor $\ell$ into $B_\xi$
equal-width bins $I_{\ell1},\ldots,I_{\ell B_\xi}$.  Boundary points
are assigned by a fixed half-open convention, with the last bin containing
the right endpoint.  Write
\[
\mathcal I_{\ell b}
=\{i:x_{i\ell}\in I_{\ell b}\},
\qquad
n_{\ell b}=|\mathcal I_{\ell b}|.
\]
Empty bins have zero mass in the histogram components below.

\paragraph{Residual-guided observation weights.}
For a nonempty bin, define
\[
\overline{|r|}_{\ell b}
=
\frac{1}{n_{\ell b}}
\sum_{i\in\mathcal I_{\ell b}}|r_i|,
\qquad
d_{\ell b}(\boldsymbol r)
=
\left(
\overline{|r|}_{\ell b}+\varepsilon_r
\right)^{1/T},
\]
where $T\geq1$.  The implementation uses
\[
\varepsilon_r
=10^{-3}\left(n^{-1}\sum_{i=1}^n|r_i|\right)+10^{-12}.
\]
Every observation within the same bin receives the same unnormalized
weight, so that
\begin{equation}
\label{eq:supp_anchor}
w_i(\boldsymbol r;\ell)
=
\frac{
d_{\ell,b_\ell(i)}(\boldsymbol r)
}{
\displaystyle
\sum_{h=1}^{B_\xi}
n_{\ell h}d_{\ell h}(\boldsymbol r)
},
\qquad i=1,\ldots,n.
\end{equation}
These weights continue to be used in the unchanged even-degree proposal.

\paragraph{Even degrees.}
For an even degree $c$, let
\[
k_c=\frac{c+2}{2}.
\]
Draw an auxiliary observation index $s$ with probability
$w_s(\boldsymbol r;\ell)$.  Conditional on $x_{s\ell}$, draw $k_c$
knots independently from $U(a_\ell,x_{s\ell})$ and $k_c$ knots
independently from $U(x_{s\ell},b_\ell)$, and order the knots on each
side.  The auxiliary anchor itself does not enter the knot vector.
Conditional on $s$, the density is
\begin{align*}
&q_{c,\ell}(\boldsymbol\xi\mid s)   \\
&=
\frac{(k_c!)^2}
{(x_{s\ell}-a_\ell)^{k_c}
 (b_\ell-x_{s\ell})^{k_c}}
\mathbf1\!\left\{
a_\ell<\xi_1<\cdots<\xi_{k_c}<x_{s\ell}
<\xi_{k_c+1}<\cdots<\xi_{2k_c}<b_\ell
\right\}.
\end{align*}
After marginalizing the auxiliary index,
\begin{equation}
\label{eq:supp_knot_even}
q_{c,\ell}(\boldsymbol\xi\mid\boldsymbol r)
=
\sum_{
s:\,\xi_{k_c}<x_{s\ell}<\xi_{k_c+1}
}
w_s(\boldsymbol r;\ell)
q_{c,\ell}(\boldsymbol\xi\mid s).
\end{equation}
The even-degree prior remains uniform on the continuous ordered-knot
domain,
\begin{equation}
\label{eq:supp_knot_prior_even}
\pi_\xi(\boldsymbol\xi\mid c,\ell)
=
\frac{(c+2)!}{(b_\ell-a_\ell)^{c+2}},
\end{equation}
and hence
\begin{equation}
\label{eq:supp_knot_ratio_even}
\frac{
\pi_\xi(\boldsymbol\xi\mid c,\ell)
}{
q_{c,\ell}(\boldsymbol\xi\mid\boldsymbol r)
}
=
\frac{
(c+2)!/(b_\ell-a_\ell)^{c+2}
}{
\displaystyle
\sum_{
s:\,\xi_{k_c}<x_{s\ell}<\xi_{k_c+1}
}
w_s(\boldsymbol r;\ell)
q_{c,\ell}(\boldsymbol\xi\mid s)
}.
\end{equation}

\paragraph{Odd-degree design-based prior.}
For an odd degree $c$, let
\[
k_c=\frac{c+1}{2},
\qquad
m_c=\frac{c+3}{2},
\qquad
A=\xi_{m_c}.
\]
The central knot $A$ is continuous.  Its prior density is
\begin{equation}
\label{eq:supp_odd_prior_center}
h_\ell(u)
=
(1-\eta_0)
\sum_{b=1}^{B_\xi}
\frac{n_{\ell b}}{n}
\frac{\mathbf1\{u\in I_{\ell b}\}}{|I_{\ell b}|}
+
\eta_0
\frac{\mathbf1\{a_\ell<u<b_\ell\}}{b_\ell-a_\ell},
\qquad 0<\eta_0\leq1.
\end{equation}
The implementation fixes $\eta_0=0.05$.  The histogram part uses only
the observed predictor locations.  The uniform component makes
$h_\ell(u)>0$ throughout the knot domain, including the portions of the
expanded domain outside the observed range.  If the observed predictor
range is degenerate, the implementation uses the uniform density on
$(a_\ell,b_\ell)$.

Conditional on $A$, the side knots have density
$p_{\mathrm{side}}(\boldsymbol\xi_{-m_c}\mid A)$ in
\eqref{eq:supp_side_density}.  Thus the odd-degree knot prior is
\begin{equation}
\label{eq:supp_knot_prior_odd}
\pi_\xi(\boldsymbol\xi\mid c,\ell,\mathbf X)
=
h_\ell(\xi_{m_c})
p_{\mathrm{side}}
(\boldsymbol\xi_{-m_c}\mid\xi_{m_c}).
\end{equation}
All $c+2$ knot coordinates are continuous.

\paragraph{Odd-degree residual-guided proposal.}
The residual-guided mass assigned to bin $b$ is
\[
\omega_{\ell b}(\boldsymbol r)
=
\frac{
n_{\ell b}d_{\ell b}(\boldsymbol r)
}{
\displaystyle
\sum_{h=1}^{B_\xi}
n_{\ell h}d_{\ell h}(\boldsymbol r)
}.
\]
This gives the histogram density
\[
\widetilde g_\ell(u\mid\boldsymbol r)
=
\sum_{b=1}^{B_\xi}
\omega_{\ell b}(\boldsymbol r)
\frac{\mathbf1\{u\in I_{\ell b}\}}{|I_{\ell b}|}.
\]
To retain full support, mix it with a uniform density on the knot domain:
\begin{equation}
\label{eq:supp_odd_proposal_center}
g_\ell(u\mid\boldsymbol r)
=
(1-\eta)
\widetilde g_\ell(u\mid\boldsymbol r)
+
\eta
\frac{\mathbf1\{a_\ell<u<b_\ell\}}{b_\ell-a_\ell},
\qquad 0<\eta\leq1.
\end{equation}
The tuning parameter $\eta$ is called \texttt{odd\_anchor\_uniform} in
the software and has default value $0.05$.  After drawing
$A\sim g_\ell(\cdot\mid\boldsymbol r)$, the side knots are drawn from
$p_{\mathrm{side}}(\cdot\mid A)$.  The resulting proposal density is
\begin{equation}
\label{eq:supp_knot_odd_q}
q_{c,\ell}(\boldsymbol\xi\mid\boldsymbol r)
=
g_\ell(\xi_{m_c}\mid\boldsymbol r)
p_{\mathrm{side}}
(\boldsymbol\xi_{-m_c}\mid\xi_{m_c}).
\end{equation}
Consequently,
\begin{equation}
\label{eq:supp_knot_ratio_odd}
\frac{
\pi_\xi(\boldsymbol\xi\mid c,\ell,\mathbf X)
}{
q_{c,\ell}(\boldsymbol\xi\mid\boldsymbol r)
}
=
\frac{h_\ell(\xi_{m_c})}
{g_\ell(\xi_{m_c}\mid\boldsymbol r)}.
\end{equation}
The prior $h_\ell$ is fixed given $\mathbf X$, whereas $g_\ell$ is
recomputed from the residual relevant to the proposed move.  Initialization
of an odd-degree factor draws its center from $h_\ell$ and its side knots
from $p_{\mathrm{side}}$.

\paragraph{Knot correction.}
For a birth move, the knot contribution for the proposed basis is
\[
\prod_{k=1}^{K_{\mathrm{new}}}
\frac{
\pi_\xi(\boldsymbol\xi_k\mid c_k,\nu_k,\mathbf X)
}{
q_{c_k,\nu_k}(\boldsymbol\xi_k\mid\boldsymbol r)
}.
\]
For an odd-degree factor this contribution is precisely
$h_{\nu_k}(\xi_{k,m_{c_k}})/
g_{\nu_k}(\xi_{k,m_{c_k}}\mid\boldsymbol r)$; the even-degree factor is
given by \eqref{eq:supp_knot_ratio_even}.  For a death move, the reciprocal
correction is evaluated using the residual under the proposed smaller
model.  For a degree--knot refresh, the correction for the proposed knot
state is divided by the correction for the current state.

\subsection{Birth and death moves}
\label{subsec:supp_bd}

\paragraph{Move probabilities.}
Let $N_{B,\max}$ denote the maximum allowed number of basis
functions. The implementation restricts $J$ to
\[
1\leq J\leq N_{B,\max}.
\]
Let $p_b$ and $p_d$ be the interior birth and death probabilities,
where
\[
p_b\geq0,
\qquad
p_d\geq0,
\qquad
p_b+p_d\leq1.
\]
The actual move probabilities are
\begin{equation}
\label{eq:supp_move_probs}
g_b(J)
=
\begin{cases}
1, & J=1,\\
p_b, & 1<J<N_{B,\max},\\
0, & J=N_{B,\max},
\end{cases}
\qquad
g_d(J)
=
\begin{cases}
0, & J=1,\\
p_d, & 1<J<N_{B,\max},\\
1, & J=N_{B,\max}.
\end{cases}
\end{equation}
Thus, a birth is forced at $J=1$, whereas a death is forced at
$J=N_{B,\max}$. At an interior state, the birth--death step is skipped
with probability $1-p_b-p_d$. The within-model backfitting and refresh
updates are performed in every iteration regardless of this choice.

The move-selection contribution to the birth acceptance ratio is
\[
\frac{g_d(J+1)}{g_b(J)}
=
\begin{cases}
p_d, & J=1,\\[4pt]
\dfrac{p_d}{p_b},
& 2\leq J\leq N_{B,\max}-2,\\[8pt]
\dfrac{1}{p_b},
& J=N_{B,\max}-1.
\end{cases}
\]
Similarly, the corresponding contribution to the death acceptance
ratio is
\[
\frac{g_b(J-1)}{g_d(J)}
=
\begin{cases}
\dfrac{1}{p_d}, & J=2,\\[8pt]
\dfrac{p_b}{p_d},
& 3\leq J\leq N_{B,\max}-1,\\[8pt]
p_b, & J=N_{B,\max}.
\end{cases}
\]
These are the reverse-to-forward probabilities of selecting the
corresponding move types.

\paragraph{Birth move.}
A birth proposes a new structural parameter
\[
\psi_{\mathrm{new}}
=
(
K_{\mathrm{new}},
\boldsymbol\nu_{\mathrm{new}},
\boldsymbol c_{\mathrm{new}},
\boldsymbol\xi_{\mathrm{new}}
).
\]
The coefficient of the proposed basis is integrated out when the
structural move is evaluated and is sampled only after the birth is
accepted.

Let
\[
\boldsymbol r
=
\boldsymbol Y
-
\beta_0\mathbf1_n
-
\sum_{j=1}^J
\beta_j\boldsymbol b_j
\]
be the current residual, and let
$\boldsymbol b_{\mathrm{new}}$ denote the proposed basis vector. Its
collapsed marginal contribution is
\[
\exp\{
\ell(
\boldsymbol b_{\mathrm{new}};
\boldsymbol r)
\},
\]
where $\ell$ is defined in
\eqref{eq:supp_collapsed_score}.

The structural prior factorizes as
\[
\begin{aligned}
\pi(
\psi_{\mathrm{new}}
\mid\boldsymbol\pi)
=\;
\pi_K(K_{\mathrm{new}})
\,
p_{\mathrm{WOR}}
(
\boldsymbol\nu_{\mathrm{new}}
\mid K_{\mathrm{new}},\boldsymbol\pi)
\times
\prod_{l=1}^{K_{\mathrm{new}}}
\pi_d(c_l)
\,
\pi_\xi(
\xi_l
\mid
c_l,\nu_l).
\end{aligned}
\]

The corresponding structural proposal factorizes as
\begin{equation}
\label{eq:supp_qbirth}
\begin{aligned}
q_{\mathrm{birth}}^{\mathrm{str}}
(
\psi_{\mathrm{new}}
\mid
\boldsymbol r,\boldsymbol\pi)
=
q_K(K_{\mathrm{new}})
q_\nu(
\boldsymbol\nu_{\mathrm{new}}
\mid
K_{\mathrm{new}},
\widetilde{\boldsymbol\pi})
q_d(
\boldsymbol c_{\mathrm{new}}
\mid K_{\mathrm{new}})
\times
q_{\mathrm{knot}}
(
\boldsymbol\xi_{\mathrm{new}}
\mid
\boldsymbol r,
\boldsymbol\nu_{\mathrm{new}},
\boldsymbol c_{\mathrm{new}}).
\end{aligned}
\end{equation}
We choose $q_K$ to equal the interaction-order prior, and the degree
proposal is the product of uniform distributions over $S$, so it also
matches the corresponding degree prior. Consequently, these factors
cancel from the structural prior--proposal ratio.

Under
\[
J\mid M,\lambda_J
\sim
\operatorname{Poi}(Me^{-\lambda_J}),
\]
the dimension-prior ratio is
\begin{equation}
\label{eq:supp_J_ratio}
\frac{
\pi(J+1\mid M,\lambda_J)
}{
\pi(J\mid M,\lambda_J)
}
=
\frac{Me^{-\lambda_J}}{J+1}.
\end{equation}

The basis functions are exchangeably labeled. In a birth move, the
newly proposed basis is assigned uniformly to one of the $J+1$ basis
labels. Equivalently, one may append the new basis and then apply a
uniform random relabeling. The resulting insertion probability
$1/(J+1)$ cancels the probability $1/(J+1)$ of selecting the same
basis in the corresponding reverse death move.

The birth acceptance ratio is therefore
\begin{equation*}
\begin{aligned}
A_{\mathrm{birth}}
=\;
\exp\{
\ell(
\boldsymbol b_{\mathrm{new}};
\boldsymbol r)
\}
\frac{Me^{-\lambda_J}}{J+1}
\times
\frac{
\pi(
\psi_{\mathrm{new}}
\mid\boldsymbol\pi)
}{
q_{\mathrm{birth}}^{\mathrm{str}}
(
\psi_{\mathrm{new}}
\mid
\boldsymbol r,\boldsymbol\pi)
}
\frac{
g_d(J+1)
}{
g_b(J)
}.
\end{aligned}
\end{equation*}
Because the interaction-order proposal equals its prior and the degree
proposal equals the degree prior, these terms cancel. The birth
acceptance ratio can therefore be written explicitly as
\begin{equation}
\label{eq:supp_birth}
\begin{aligned}
A_{\mathrm{birth}}
=\;
\exp\left\{
\ell(
\boldsymbol b_{\mathrm{new}};
\boldsymbol r)
\right\}
\frac{Me^{-\lambda_J}}{J+1}
\times
\frac{
p_{\mathrm{WOR}}
\left(
\boldsymbol\nu_{\mathrm{new}}
\mid
K_{\mathrm{new}},
\boldsymbol\pi
\right)
}{
p_{\mathrm{WOR}}
\left(
\boldsymbol\nu_{\mathrm{new}}
\mid
K_{\mathrm{new}},
\widetilde{\boldsymbol\pi}(\boldsymbol r)
\right)
}
\times
\prod_{l=1}^{K_{\mathrm{new}}}
\frac{
\pi_\xi
\left(
\boldsymbol\xi_l^{\mathrm{new}}
\mid
c_l^{\mathrm{new}},
\nu_l^{\mathrm{new}}
\right)
}{
q_{c_l^{\mathrm{new}},\nu_l^{\mathrm{new}}}
\left(
\boldsymbol\xi_l^{\mathrm{new}}
\mid
\boldsymbol r
\right)
}
\frac{
g_d(J+1)
}{
g_b(J)
}.
\end{aligned}
\end{equation}
The proposed birth is accepted with probability
\[
\alpha_{\mathrm{birth}}
=
\min\{1,A_{\mathrm{birth}}\}.
\]

The acceptance ratio contains the exact residual-guided corrections for
active-predictor and knot selection.  In particular, every odd-degree
factor contributes
\[
\frac{h_\ell(A)}{g_\ell(A\mid\boldsymbol r)}
\]
to the structural prior--proposal ratio.  No coefficient-proposal density
appears because the coefficient is integrated out before the acceptance
decision.

If accepted, the new coefficient is drawn from
$N(m,v^2)$ using \eqref{eq:supp_collapsed_moments} with
$\boldsymbol b=\boldsymbol b_{\mathrm{new}}$ and the current residual
$\boldsymbol r$.

\paragraph{Death move.}
A death move selects
\[
j^\ast
\sim
\operatorname{Uniform}\{1,\ldots,J\}
\]
and proposes to remove the corresponding basis function. Define the
residual under the proposed smaller model as
\begin{equation}
\label{eq:supp_death_resid}
\boldsymbol r_{-j^\ast}
=
\boldsymbol Y
-
\beta_0\mathbf1_n
-
\sum_{j\ne j^\ast}
\beta_j\boldsymbol b_j.
\end{equation}

All quantities needed to evaluate the reverse birth proposal,
including the residual relevance scores, tilted variable-selection
probabilities, and residual-guided knot proposal densities, are
computed using $\boldsymbol r_{-j^\ast}$.

The death acceptance ratio is
\begin{equation}
\label{eq:supp_death}
\begin{aligned}
A_{\mathrm{death}}
=&\;
\exp\{
-\ell(
\boldsymbol b_{j^\ast};
\boldsymbol r_{-j^\ast})
\}
\frac{J}{Me^{-\lambda_J}}
\times
\frac{
q_{\mathrm{birth}}^{\mathrm{str}}
(
\psi_{j^\ast}
\mid
\boldsymbol r_{-j^\ast},
\boldsymbol\pi)
}{
\pi(
\psi_{j^\ast}
\mid
\boldsymbol\pi)
}
\frac{
g_b(J-1)
}{
g_d(J)
} \\
=&\;
\exp\left\{
-\ell(
\boldsymbol b_{j^\ast};
\boldsymbol r_{-j^\ast})
\right\}
\frac{J}{Me^{-\lambda_J}}
\times
\frac{
p_{\mathrm{WOR}}
\left(
\boldsymbol\nu^{(j^\ast)}
\mid
K_{j^\ast},
\widetilde{\boldsymbol\pi}_{-j^\ast}
\right)
}{
p_{\mathrm{WOR}}
\left(
\boldsymbol\nu^{(j^\ast)}
\mid
K_{j^\ast},
\boldsymbol\pi
\right)
}
\\
&\times
\prod_{l=1}^{K_{j^\ast}}
\frac{
q_{c_l^{(j^\ast)},\nu_l^{(j^\ast)}}
\left(
\boldsymbol\xi_l^{(j^\ast)}
\mid
\boldsymbol r_{-j^\ast}
\right)
}{
\pi_\xi
\left(
\boldsymbol\xi_l^{(j^\ast)}
\mid
c_l^{(j^\ast)},
\nu_l^{(j^\ast)}
\right)
}
\frac{
g_b(J-1)
}{
g_d(J)
},
\end{aligned}
\end{equation}
where $
\widetilde{\boldsymbol\pi}_{-j^\ast}
=
\widetilde{\boldsymbol\pi}
\left(
\boldsymbol r_{-j^\ast}
\right)$.
The proposed death is accepted with probability
\[
\alpha_{\mathrm{death}}
=
\min\{1,A_{\mathrm{death}}\}.
\]
Equation~\eqref{eq:supp_death} is the reciprocal of the corresponding
reverse-birth ratio evaluated at the proposed smaller model.

Thus both the reverse variable-selection probabilities and the reverse
knot proposal are recomputed from $\boldsymbol r_{-j^\ast}$.  An odd-degree
factor in the deleted basis contributes the reciprocal correction
\[
\frac{g_\ell(A\mid\boldsymbol r_{-j^\ast})}{h_\ell(A)}.
\]

\subsection{Global parameter updates}
\label{subsec:supp_global}

\paragraph{Error variance.}
Under
\[
\sigma^2
\sim
\mathrm{IG}
\left(
\frac r2,
\frac{rR}{2}
\right),
\]
the full conditional distribution is
\begin{equation}
\label{eq:supp_sigma}
\sigma^2\mid\mathrm{rest}
\sim
\mathrm{IG}
\left(
\frac{r+n}{2},
\frac{
rR+
\|
\boldsymbol Y-\boldsymbol\mu_J
\|^2
}{2}
\right).
\end{equation}

\paragraph{Total mass.}
Write
\[
c_J=e^{-\lambda_J},
\qquad
\phi_M^2=\frac{\tau^2}{c_JM}.
\]
The terms of the joint posterior that depend on $M$ are
\[
\pi(M)
\propto
M^{a_\gamma-1}e^{-b_\gamma M},
\]
\[
\pi(J\mid M)
\propto
M^J e^{-c_JM},
\]
and
\[
\prod_{j=1}^J
N
\left(
\beta_j;
0,
\frac{\tau^2}{c_JM}
\right)
\propto
M^{J/2}
\exp
\left\{
-\frac{c_JM}{2\tau^2}
\sum_{j=1}^J\beta_j^2
\right\}.
\]
Therefore,
\begin{equation}
\label{eq:supp_M}
M\mid\mathrm{rest}
\sim
\mathrm{Ga}
\left(
a_\gamma+\frac{3J}{2},
\;
b_\gamma+
e^{-\lambda_J}
\left[
1+
\frac{1}{2\tau^2}
\sum_{j=1}^J
\beta_j^2
\right]
\right),
\end{equation}
where the Gamma distribution is parameterized by shape and rate.

\paragraph{Variable-selection probabilities.}
Because the active predictors within each basis are sampled without
replacement, the full conditional distribution of
$\boldsymbol\pi=(\pi_1,\ldots,\pi_p)$ is not Dirichlet. Let
\[
n_\ell
=
\sum_{j=1}^J
\sum_{m=1}^{K_j}
\mathbf1
\{
\nu_m^{(j)}=\ell
\}
\]
denote the number of occurrences of predictor $\ell$ among the active
predictors. Up to a normalizing constant, the target density is
\begin{equation}
\label{eq:supp_pi_target}
h(\boldsymbol\pi)
=
\left[
\prod_{\ell=1}^p
\pi_\ell^{\alpha/p-1}
\right]
\prod_{j=1}^J
p_{\mathrm{WOR}}
(
\boldsymbol\nu^{(j)}
\mid
K_j,\boldsymbol\pi).
\end{equation}

We use the DART-style Dirichlet independence proposal
\citep{linero2018dart}
\[
q_\pi(
\boldsymbol\pi^\star)
=
\operatorname{Dir}
\left(
\frac{\alpha}{p}+n_1,
\ldots,
\frac{\alpha}{p}+n_p
\right).
\]
The proposal is accepted with probability
$\min\{1,A_\pi\}$, where
\[
\log A_\pi
=
\log h(\boldsymbol\pi^\star)
-
\log h(\boldsymbol\pi)
+
\log q_\pi(\boldsymbol\pi)
-
\log q_\pi(\boldsymbol\pi^\star).
\]

During the initial variable-selection warm-up period,
$\boldsymbol\pi$ is held fixed at the uniform vector
$\mathbf1_p/p$. This prevents premature concentration on a small
subset of predictors before the sampler has sufficiently explored the
predictor space.

\clearpage
\subsection{Overall posterior sampling algorithm}
\label{subsec:supp_alg}

Algorithm~\ref{alg:mlabs_supp} summarizes one MCMC sweep.

\begin{algorithm}[H]
\caption{One MCMC sweep for MLABS}
\label{alg:mlabs_supp}

\begin{adjustbox}{
    max totalsize={\linewidth}{0.85\textheight},
    center
}
\begin{minipage}{\linewidth}

\begin{algorithmic}[1]
\Require Current state and fixed tuning parameters

\Statex \textbf{Phase 1: Within-model backfitting}
\For{$j=1,\ldots,J$}
    \State Form $\boldsymbol r^{(j)}$ and sample $\beta_j$ from
    \eqref{eq:supp_beta_fc}
    \State Update all even-degree knots and the side knots of odd-degree
    factors using \eqref{eq:supp_local_knot}
    \State For each odd-degree factor, jointly propose a continuous center
    and new side knots using \eqref{eq:supp_anchor_update}
\EndFor

\Statex \textbf{Phase 2: Joint degree--knot refresh}
\State Select $\min(n_{\mathrm{ref}},J)$ distinct bases uniformly without
replacement
\For{each selected basis $j$}
    \State Form $\boldsymbol r^{(j)}$ and propose
    $(\boldsymbol c^{(j)},\boldsymbol\xi^{(j)})
    \rightarrow
    (\boldsymbol c^{(j)\prime},\boldsymbol\xi^{(j)\prime})$
    \State Accept with probability $\min\{1,A_{\mathrm{ref}}\}$ from
    \eqref{eq:supp_refresh}
    \If{accepted}
        \State Recompute $\boldsymbol b_j$ and sample $\beta_j$ from
        \eqref{eq:supp_collapsed_moments}
    \EndIf
\EndFor

\Statex \textbf{Phase 3: Birth/death update}
\If{$J=1$}
    \State Set the move type to birth
\ElsIf{$J=N_{B,\max}$}
    \State Set the move type to death
\Else
    \State Draw birth, death, or skip with probabilities
    $p_b$, $p_d$, and $1-p_b-p_d$
\EndIf
\If{birth}
    \State Form the current residual $\boldsymbol r$
    \State Draw $K_{\mathrm{new}}$, compute $\widetilde{\boldsymbol\pi}$,
    and draw active predictors without replacement
    \State Draw degrees from their prior and knots from
    Section~\ref{subsec:supp_knot}
    \State Accept with probability $\min\{1,A_{\mathrm{birth}}\}$ from
    \eqref{eq:supp_birth}
    \If{accepted}
        \State Draw the new coefficient from
        \eqref{eq:supp_collapsed_moments} and randomly assign its label
    \EndIf
\ElsIf{death}
    \State Draw $j^\ast$ uniformly, form $\boldsymbol r_{-j^\ast}$, and
    evaluate the reverse birth proposal using this residual
    \State Accept removal with probability
    $\min\{1,A_{\mathrm{death}}\}$ from \eqref{eq:supp_death}
\EndIf

\Statex \textbf{Phase 4: Global updates.} Update $\boldsymbol\mu_J$;
sample $\sigma^2$ and $M$ from \eqref{eq:supp_sigma} and
\eqref{eq:supp_M}; after warm-up, update $\boldsymbol\pi$ using
\eqref{eq:supp_pi_target}
\end{algorithmic}
\end{minipage}
\end{adjustbox}
\end{algorithm}
\clearpage

\newpage


\section{Proofs of the Theoretical Results}
\label{sec:supp_theory}

This part provides the supporting lemmas and proofs of all results stated
in Section 5 of the main text.

\subsection{Proof of Theorem 1}
\label{subsec:supp_proof_ua}

\begin{lemma}[Rectangle indicators from degree-zero B-splines]
\label{lem:rectangle_indicator}
Let
\[
R=\prod_{i=1}^p[a_i,b_i),
\qquad 0\leq a_i<b_i\leq1.
\]
For each interval $[a_i,b_i)$, define the local knot vector
\[
\boldsymbol{\xi}_{a_i,b_i}=(a_i,b_i).
\]
Under the standard half-open support convention,
\[
B_0(x_i;\boldsymbol{\xi}_{a_i,b_i})
=
\mathbf{1}_{[a_i,b_i)}(x_i).
\]
Then
\[
\mathbf{1}_R(\mathbf{x})
=
\prod_{i=1}^p
B_0(x_i;\boldsymbol{\xi}_{a_i,b_i})
\]
for every $\mathbf{x}\in[0,1]^p$.
\end{lemma}

\begin{proof}
By the definition of a Cartesian product,
\[
\mathbf{1}_R(\mathbf{x})
=
\prod_{i=1}^p\mathbf{1}_{[a_i,b_i)}(x_i).
\]
Substituting the degree-zero B-spline representation of each
interval indicator gives
\[
\mathbf{1}_R(\mathbf{x})
=
\prod_{i=1}^p
B_0(x_i;\boldsymbol{\xi}_{a_i,b_i}).
\]
Under other endpoint conventions, the same identity holds almost
everywhere, which is sufficient for the $L^2$ statements below.
\end{proof}

\begin{lemma}[Density of rectangle simple functions]
\label{lem:simple_density}
For every $f\in L^2([0,1]^p)$ and every $\varepsilon>0$,
there exist half-open axis-aligned rectangles
\[
R_1,\ldots,R_M\subset[0,1]^p
\]
and coefficients $c_1,\ldots,c_M\in\mathbb{R}$ such that
\[
\left\|
f-\sum_{m=1}^M c_m\mathbf{1}_{R_m}
\right\|_{L^2([0,1]^p)}
<\varepsilon.
\]
Equivalently, the class
\[
\mathcal{R}
=
\left\{
\sum_{m=1}^M c_m\mathbf{1}_{R_m}
:
\begin{array}{l}
M\in\mathbb{N},\quad c_m\in\mathbb{R},\\
R_m=\prod_{i=1}^p[a_{mi},b_{mi}),\\
0\leq a_{mi}<b_{mi}\leq1
\end{array}
\right\}
\]
is dense in $L^2([0,1]^p)$.
\end{lemma}

\begin{proof}
Let $\lambda$ denote Lebesgue measure restricted to $[0,1]^p$.
Since simple functions are dense in $L^2([0,1]^p)$, there exists
\[
s(\mathbf{x})
=
\sum_{m=1}^{M_0}\gamma_m\mathbf{1}_{A_m}(\mathbf{x}),
\qquad
A_m\subset[0,1]^p \text{ measurable},
\]
such that
\[
\|f-s\|_{L^2([0,1]^p)}<\frac{\varepsilon}{2}.
\]
Choose $\varepsilon_m>0$ so that
\[
\sum_{m=1}^{M_0}|\gamma_m|\varepsilon_m^{1/2}
<\frac{\varepsilon}{2}.
\]

By regularity of $\lambda$, each $A_m$ can be approximated
in symmetric-difference measure by a finite union of closed
axis-aligned rectangles. To see this, choose a compact set
$K_m$ and a relatively open set $O_m$ in $[0,1]^p$ such that
\[
K_m\subset A_m\subset O_m,
\qquad
\lambda(O_m\setminus K_m)<\varepsilon_m.
\]
Each point of $K_m$ has a relative neighborhood contained in
a closed axis-aligned rectangle lying in $O_m$.
Compactness gives finitely many such rectangles
$Q_{m1},\ldots,Q_{mN_m}$ whose union $B_m$ satisfies
\[
K_m\subset B_m
:=
\bigcup_{j=1}^{N_m}Q_{mj}
\subset O_m.
\]
If $K_m$ is empty, we may take $B_m=\varnothing$.
Let
\[
A_m\triangle B_m
:=
(A_m\setminus B_m)\cup(B_m\setminus A_m)
\]
denote the symmetric difference of $A_m$ and $B_m$.
Since $K_m\subset A_m,B_m\subset O_m$, we have
\[
A_m\triangle B_m\subset O_m\setminus K_m,
\]
and hence
\[
\lambda(A_m\triangle B_m)
\leq \lambda(O_m\setminus K_m)
<\varepsilon_m.
\]

Define
\[
\widetilde{s}(\mathbf{x})
=
\sum_{m=1}^{M_0}\gamma_m\mathbf{1}_{B_m}(\mathbf{x}).
\]
For each nonempty union $B_m$, inclusion--exclusion gives
\[
\mathbf{1}_{B_m}
=
\sum_{\substack{
I\subseteq\{1,\ldots,N_m\}\\
I\neq\varnothing
}}
(-1)^{|I|+1}
\mathbf{1}_{\bigcap_{j\in I}Q_{mj}}.
\]
Every finite intersection $\bigcap_{j\in I}Q_{mj}$ is either
a closed axis-aligned rectangle or empty. Thus
$\widetilde{s}$ is a finite linear combination of closed
rectangle indicators.

Discard any degenerate rectangles, which have Lebesgue measure
zero, and replace each remaining closed rectangle
$\prod_{i=1}^p[a_i,b_i]$ by
$\prod_{i=1}^p[a_i,b_i)$.
These replacements change the indicators only on sets of
Lebesgue measure zero. Consequently, there exist half-open
rectangles $R_1,\ldots,R_M$ and coefficients
$c_1,\ldots,c_M$ such that
\[
\widetilde{s}
=
\sum_{m=1}^M c_m\mathbf{1}_{R_m}
\quad\text{almost everywhere}.
\]
If no terms remain, use any nondegenerate half-open rectangle
with coefficient zero.

Since
\[
\left|\mathbf{1}_{A_m}-\mathbf{1}_{B_m}\right|^2
=
\mathbf{1}_{A_m\triangle B_m},
\]
we have
\[
\|\mathbf{1}_{A_m}-\mathbf{1}_{B_m}\|_{L^2([0,1]^p)}
=
\lambda(A_m\triangle B_m)^{1/2}.
\]
The triangle inequality therefore yields
\begin{align*}
\left\|
s-\sum_{m=1}^M c_m\mathbf{1}_{R_m}
\right\|_{L^2([0,1]^p)}
&=
\|s-\widetilde{s}\|_{L^2([0,1]^p)}
\\
&\leq
\sum_{m=1}^{M_0}
|\gamma_m|
\|\mathbf{1}_{A_m}-\mathbf{1}_{B_m}\|_{L^2([0,1]^p)}
\\
&=
\sum_{m=1}^{M_0}
|\gamma_m|\lambda(A_m\triangle B_m)^{1/2}
\\
&\leq
\sum_{m=1}^{M_0}
|\gamma_m|\varepsilon_m^{1/2}
<
\frac{\varepsilon}{2}.
\end{align*}
Finally,
\begin{align*}
\left\|
f-\sum_{m=1}^M c_m\mathbf{1}_{R_m}
\right\|_{L^2([0,1]^p)}
&\leq
\|f-s\|_{L^2([0,1]^p)}
+
\left\|
s-\sum_{m=1}^M c_m\mathbf{1}_{R_m}
\right\|_{L^2([0,1]^p)}
\\
&<
\varepsilon.
\end{align*}
This proves the claim.
\end{proof}

\begin{proof}
Let
\[
R=\prod_{i=1}^p[a_i,b_i),
\qquad 0\leq a_i<b_i\leq1.
\]
By Lemma~\ref{lem:rectangle_indicator},
\[
\mathbf{1}_R(\mathbf{x})
=
\prod_{i=1}^p
B_0(x_i;\boldsymbol{\xi}_{a_i,b_i})
\]
almost everywhere on $[0,1]^p$.

Since $a_i<b_i$, the two knots are distinct and correctly ordered.
Moreover, the assumption that each predictor's knot domain
contains $[0,1]$ in its interior ensures that these knots are
admissible, including when $a_i=0$ or $b_i=1$.
Since $0\in S$ and $K_{\max}\geq p$, the product uses $p$
distinct predictors and is a valid MLABS atom.
Taking its coefficient to be one and the intercept to be zero
gives
\[
\mathbf{1}_R\in\mathcal{F}(1,K_{\max},S),
\]
where functions are identified up to equality almost everywhere.

Now fix $f\in L^2([0,1]^p)$ and $\varepsilon>0$.
By Lemma~\ref{lem:simple_density}, there exist half-open
rectangles $R_1,\ldots,R_M$ and real coefficients
$c_1,\ldots,c_M$ such that
\[
g(\mathbf{x})
=
\sum_{m=1}^M c_m\mathbf{1}_{R_m}(\mathbf{x})
\]
satisfies
\[
\|f-g\|_{L^2([0,1]^p)}<\varepsilon.
\]
Replacing each rectangle indicator by its product of
degree-zero B-splines expresses $g$ as an MLABS expansion
with at most $M$ terms and intercept zero. Hence
\[
g\in\mathcal{F}(M,K_{\max},S)
\subseteq\mathcal{F}_\infty(K_{\max},S).
\]
Since $f$ and $\varepsilon$ were arbitrary,
$\mathcal{F}_\infty(K_{\max},S)$ is dense in $L^2([0,1]^p)$.
\end{proof}

\subsection{Proof of Theorem 2}

\begin{definition}[Modulus of smoothness and Besov space]
\label{def:besov}
Let $\Omega=[0,1]^d$ and $0<a\leq\infty$.
For an integer $r\geq 1$ and a step $h\in\mathbb{R}^d$, the $r$th-order
forward difference of $f$ is
\[
\Delta_h^r f(\mathbf{x})
=
\sum_{i=0}^{r}\binom{r}{i}(-1)^{r-i}f(\mathbf{x}+ih),
\qquad
\mathbf{x}\in\Omega_{rh}:=\{\mathbf{x}\in\Omega:\mathbf{x}+ih\in\Omega,\ i=0,\dots,r\},
\]
and the $r$th-order $L^a$ modulus of smoothness of $f\in L^a(\Omega)$ is
\[
\omega_r(f,t)_a
:=
\sup_{0<|h|\leq t}
\bigl\|\Delta_h^r f\bigr\|_{L^a(\Omega_{rh})},
\qquad t>0 .
\]
For a smoothness index $s>0$, integrability $0<a\leq\infty$, and
fine-scale index $0<q\leq\infty$, fix any integer $r>s$ (in this paper
$r=k+1$, where $k$ is the spline degree).
The \emph{Besov space} $B^s_{a,q}(\Omega)$ is the set of all
$f\in L^a(\Omega)$ for which the seminorm
\[
|f|_{B^s_{a,q}(\Omega)}
:=
\begin{cases}
\displaystyle
\left(
\int_0^1
\bigl(t^{-s}\,\omega_r(f,t)_a\bigr)^{q}
\,\frac{dt}{t}
\right)^{1/q},
& 0<q<\infty,\\[1.1em]
\displaystyle
\sup_{0<t\leq 1} t^{-s}\,\omega_r(f,t)_a,
& q=\infty,
\end{cases}
\]
is finite, equipped with the (quasi-)norm
\[
\|f\|_{B^s_{a,q}(\Omega)}
:=
\|f\|_{L^a(\Omega)}
+
|f|_{B^s_{a,q}(\Omega)} .
\]
The indices have the following interpretation: $s$ controls smoothness,
$a$ controls integrability, and $q$ controls the fine-scale summability of its multiscale coefficients
\end{definition}

Throughout the following results, $A\lesssim B$ means
$A\leq CB$ for a constant $C>0$ independent of the function
being approximated and the approximation resolution.

\begin{lemma}[Besov embedding into an $L^2$ scale]
\label{lem:besov_embedding}
Let $d\geq1$ and
\[
f\in B^s_{a,q}([0,1]^d),
\qquad
s>0,
\qquad
0<a,q\leq\infty.
\]
Define
\[
\rho
=
s-d\left(\frac{1}{a}-\frac{1}{2}\right)_+,
\qquad
(x)_+=\max\{x,0\},
\]
and assume that $\rho>0$. Then
\[
B^s_{a,q}([0,1]^d)
\hookrightarrow
B^\rho_{2,\infty}([0,1]^d).
\]
In particular,
\[
\|f\|_{B^\rho_{2,\infty}([0,1]^d)}
\lesssim
\|f\|_{B^s_{a,q}([0,1]^d)}.
\]
\end{lemma}

\begin{proof}
If $a<2$, then
\[
\rho
=
s-d\left(\frac{1}{a}-\frac{1}{2}\right),
\]
so that
\[
s-\frac{d}{a}
=
\rho-\frac{d}{2}.
\]
The standard critical Besov embedding therefore gives
\[
B^s_{a,q}([0,1]^d)
\hookrightarrow
B^\rho_{2,\infty}([0,1]^d),
\]
since $a<2$ and $q\leq\infty$. If $a\geq2$, then $\rho=s$. On the bounded domain $[0,1]^d$,
the standard Besov embedding gives
\[
B^s_{a,q}([0,1]^d)
\hookrightarrow
B^s_{2,q}([0,1]^d)
\hookrightarrow
B^s_{2,\infty}([0,1]^d).
\]
Thus, in either case,
\[
B^s_{a,q}([0,1]^d)
\hookrightarrow
B^\rho_{2,\infty}([0,1]^d),
\]
which proves the result.
\end{proof}

\begin{lemma}[Admissible basis for a simple-knot spline space]
\label{lem:supp_spline_basis}
Let $k\geq0$ and
$\Delta_N=\{0=t_0<t_1<\cdots<t_N<t_{N+1}=1\}$.
Let $\mathcal S_k(\Delta_N)$ be the degree-at-most-$k$ spline
space with $C^{k-1}$ continuity at interior knots when $k\geq1$,
and no continuity requirement when $k=0$.
On any knot domain $(a,b)$ with $a<0<1<b$, this space has a basis
of $N+k+1$ degree-$k$ B-splines whose local knots are strictly
increasing in $(a,b)$. For $d\geq1$, define
\[
\mathcal S_{k,d}(\Delta_N)
:=
\mathcal S_k(\Delta_N)^{\otimes d}.
\]
Then every element of $\mathcal S_{k,d}(\Delta_N)$
admits an expansion in at most $(N+k+1)^d$
tensor-product B-splines, with each factor using
a strictly increasing local knot vector within
the corresponding coordinate's knot domain.
\end{lemma}

\begin{proof}
For $k\geq1$, choose distinct exterior knots
\[
a<t_{-k}<\cdots<t_{-1}<0,
\qquad
1<t_{N+2}<\cdots<t_{N+k+1}<b.
\]
The resulting global sequence has $N+2k+2$ knots. Its local
vectors $(t_i,\ldots,t_{i+k+1})$, $i=-k,\ldots,N$, give
$N+k+1$ degree-$k$ B-splines. The standard B-spline basis theorem
on the basic interval $[t_0,t_{N+1}]=[0,1]$ states that their
restrictions form a basis of $\mathcal S_k(\Delta_N)$;
see \citet{schumaker2007spline}. In particular, no boundary
conditions are imposed by this extension. The dimension also follows
from $k+1$ polynomial coefficients on the first interval and one
additional degree of freedom at each of the $N$ simple interior knots.
For $k=0$, the $N+1$ interval indicators already form the required
basis. Taking products of the $N+k+1$ univariate basis functions
in each of the $d$ coordinates gives a basis of
$\mathcal S_{k,d}(\Delta_N)$ with $(N+k+1)^d$ elements.
\end{proof}

\begin{lemma}[Approximation of $L^2$ Besov functions]
\label{lem:besov_spline}
Let $k\geq0$, $d\geq1$, $0<\alpha<k+1$, and
$f\in B^\alpha_{2,\infty}([0,1]^d)$.
Let $\Delta_N$ be a quasi-uniform partition with $N\geq1$
interior knots and maximum mesh size $h\asymp N^{-1}$.
Under the knot-domain assumptions of
Lemma~\ref{lem:supp_spline_basis}, there exists
$g_N\in\mathcal S_{k,d}(\Delta_N)$ such that
\[
\|f-g_N\|_{L^2([0,1]^d)}
\leq
C N^{-\alpha}
\|f\|_{B^\alpha_{2,\infty}([0,1]^d)}.
\]
Moreover, $g_N$ admits an expansion in at most
$(N+k+1)^d$ admissible tensor-product B-splines.
The constant $C$ depends only on $d$, $k$, $\alpha$,
and the mesh constants.
\end{lemma}

\begin{proof}
Set $r=k+1$. For each coordinate $i$, let $P_i$ be the univariate
$L^2$-orthogonal projection onto
$\mathcal S_k(\Delta_{N,i})$, applied in coordinate $i$.
The operators $P_i$ commute because they act on different
coordinates. Their product
\[
P=P_1\cdots P_d
\]
is the $L^2$-orthogonal projection onto
$\mathcal S_{k,d}(\Delta_N)$. Since each $P_i$ is an orthogonal projection,
\[
\|P_iw\|_2\leq\|w\|_2,
\qquad
\|Pw\|_2\leq\|w\|_2,
\qquad w\in L^2([0,1]^d).
\]

Let $v\in H^r([0,1]^d)=W^{r,2}([0,1]^d)$, the Sobolev
space with square-integrable weak derivatives of total
order at most $r$. We use the norm
\[
\|v\|_{H^r}^2
=
\sum_{|\beta|\leq r}
\|D^\beta v\|_{L^2([0,1]^d)}^2,
\]
where $\beta$ is a multi-index and $D^\beta v$ denotes
the corresponding weak derivative.
Fix $v\in H^r([0,1]^d)$ and a coordinate $i$.
For fixed values $\mathbf x_{-i}$ of the remaining
coordinates, define the univariate function
\[
v_{\mathbf x_{-i}}(t)
=
v(x_1,\ldots,x_{i-1},t,x_{i+1},\ldots,x_d).
\]
The Sobolev slicing property and Fubini's theorem imply
that $v_{\mathbf x_{-i}}\in H^r(0,1)$ for almost every
$\mathbf x_{-i}$, and its $r$th weak derivative is
$(\partial_i^r v)(t,\mathbf x_{-i})$.
We apply \citet[Theorem~1]{sande2019sharp} to these univariate
functions and integrate the squared estimates over
$\mathbf x_{-i}$. 
It follows that
\[
\|(I-P_i)v\|_2
\leq
\left(\frac{h}{\pi}\right)^r
\|\partial_i^r v\|_2.
\]
The degree condition $k\geq r-1$ holds with equality.

The identity
\[
I-P
=
\sum_{i=1}^d
P_1\cdots P_{i-1}(I-P_i),
\]
with the empty product interpreted as the identity,
and the contraction property give
\[
\begin{aligned}
\|v-Pv\|_2
&\leq
\sum_{i=1}^d\|(I-P_i)v\|_2\\
&\leq
\left(\frac{h}{\pi}\right)^r
\sum_{i=1}^d\|\partial_i^r v\|_2\\
&\leq
C_d h^r\|v\|_{H^r}.
\end{aligned}
\]
Therefore, for every $v\in H^r$,
\[
\begin{aligned}
\|f-Pf\|_2
&\leq
\|f-v\|_2+\|v-Pv\|_2+\|P(v-f)\|_2\\
&\leq
2\|f-v\|_2+C_d h^r\|v\|_{H^r}.
\end{aligned}
\]

Define the $K$-functional by
\[
K(t,f;L^2,H^r)
=
\inf_{v\in H^r}
\left\{
\|f-v\|_2+t\|v\|_{H^r}
\right\}.
\]
The preceding estimate implies
\[
\|f-Pf\|_2
\leq
C K(h^r,f;L^2,H^r).
\]
By the $K$-functional characterization of real interpolation,
\[
\|f\|_{(L^2,H^r)_{\alpha/r,\infty}}
\asymp
\sup_{t>0}
t^{-\alpha/r}K(t,f;L^2,H^r).
\]
For $0<\alpha<r$, the real-interpolation space
$(L^2,H^r)_{\alpha/r,\infty}$ coincides with
$B^\alpha_{2,\infty}([0,1]^d)$, with equivalent norms;
see \citet{triebel1983theory}.
Therefore,
\[
\sup_{t>0}
t^{-\alpha/r}K(t,f;L^2,H^r)
\leq
C\|f\|_{B^\alpha_{2,\infty}([0,1]^d)}.
\]
Hence, for every $t>0$,
\[
K(t,f;L^2,H^r)
\leq
C t^{\alpha/r}
\|f\|_{B^\alpha_{2,\infty}([0,1]^d)}.
\]
We apply this inequality with $t=h^r$ and obtain
\[
\|f-Pf\|_2
\leq
C K(h^r,f;L^2,H^r)
\leq
C h^\alpha
\|f\|_{B^\alpha_{2,\infty}([0,1]^d)}.
\]
Since $h\asymp N^{-1}$, the choice $g_N=Pf$ gives
\[
\|f-g_N\|_2
\leq
C N^{-\alpha}
\|f\|_{B^\alpha_{2,\infty}([0,1]^d)}.
\]
Finally, Lemma~\ref{lem:supp_spline_basis} provides
a representation of $g_N$ with at most $(N+k+1)^d$
admissible tensor-product B-splines.
\end{proof}

\begin{lemma}[Componentwise Besov spline approximation]
\label{lem:component_besov}
Let $k\geq0$ and $d\geq1$ be integers. Suppose that
\[
f\in B^s_{a,q}([0,1]^d),
\qquad
0<s<k+1,
\qquad
0<a,q\leq\infty.
\]
Define
\[
\rho
=
s-d\left(\frac1a-\frac12\right)_+,
\]
and assume $\rho>0$.
For a quasi-uniform partition $\Delta_N$ with $N\geq1$
interior knots, there exists
$g_N\in\mathcal S_{k,d}(\Delta_N)$ such that
\[
\|f-g_N\|_{L^2([0,1]^d)}
\leq
C N^{-\rho}
\|f\|_{B^s_{a,q}([0,1]^d)}.
\]
Under the knot-domain assumptions of
Lemma~\ref{lem:supp_spline_basis}, $g_N$ admits an expansion
in at most $(N+k+1)^d$ admissible tensor-product B-splines.
The constant $C$ depends only on $d$, $k$, $s$, $a$, $q$,
and the mesh constants, and is independent of $f$ and $N$.
\end{lemma}

\begin{proof}
Lemma~\ref{lem:besov_embedding} gives
\[
\|f\|_{B^\rho_{2,\infty}([0,1]^d)}
\leq
C\|f\|_{B^s_{a,q}([0,1]^d)}.
\]
Since $0<\rho\leq s<k+1$,
Lemma~\ref{lem:besov_spline} applies with $\alpha=\rho$.
Thus there exists $g_N\in\mathcal S_{k,d}(\Delta_N)$
such that
\[
\begin{aligned}
\|f-g_N\|_{L^2([0,1]^d)}
&\leq
C N^{-\rho}
\|f\|_{B^\rho_{2,\infty}([0,1]^d)}\\
&\leq
C N^{-\rho}
\|f\|_{B^s_{a,q}([0,1]^d)}.
\end{aligned}
\]
Lemma~\ref{lem:supp_spline_basis} gives the admissible
representation and the bound on the number of basis terms.
\end{proof}

\begin{proof}[Proof of Theorem 2]
If $\mathcal U=\varnothing$, choose $g=\mu$ and $J=0$.
Both bounds then hold exactly.
Suppose that $\mathcal U\neq\varnothing$, and fix an
integer $N\geq1$.

For each $u\in\mathcal U$, Lemma~\ref{lem:component_besov}
with degree $k$ gives a spline $g_u$ such that
\[
\|f_u-g_u\|_{L^2([0,1]^{d_u})}
\leq
C_u N^{-\rho_u}
\|f_u\|_{B^{s_u}_{a_u,q_u}([0,1]^{d_u})}.
\]
Its expansion contains at most $(N+k+1)^{d_u}$
admissible tensor-product B-splines.
Each term uses the distinct predictors in $u$ and degree
$k\in S$. Since $d_u=|u|\leq K_{\max}$ by the assumed
ANOVA structure, each term is an admissible MLABS
basis function.

Define
\[
g(\mathbf x)
=
\mu+\sum_{u\in\mathcal U}g_u(\mathbf x_u).
\]
The combined expansion satisfies
\[
g\in\mathcal F(J,K_{\max},S),
\qquad
J\leq
\sum_{u\in\mathcal U}(N+k+1)^{d_u}.
\]

For any $h_u\in L^2([0,1]^{d_u})$, the function
$h_u(\mathbf x_u)$ on $[0,1]^p$ does not depend on
the coordinates outside $u$. Fubini's theorem and
the unit volume of those coordinates imply
\begin{equation}
\label{eq:supp_lift}
\|h_u(\mathbf x_u)\|_{L^2([0,1]^p)}^2
=
\int_{[0,1]^{d_u}}
|h_u(\mathbf x_u)|^2\,d\mathbf x_u
=
\|h_u\|_{L^2([0,1]^{d_u})}^2.
\end{equation}
Therefore, the triangle inequality gives
\[
\begin{aligned}
\|f-g\|_{L^2([0,1]^p)}
&\leq
\sum_{u\in\mathcal U}
\|f_u-g_u\|_{L^2([0,1]^{d_u})}\\
&\leq
C\sum_{u\in\mathcal U}
N^{-\rho_u}
\|f_u\|_{B^{s_u}_{a_u,q_u}([0,1]^{d_u})},
\end{aligned}
\]
where $C=\max_{u\in\mathcal U}C_u$.
\end{proof}

\subsection{Proof of Corollary 3}

\begin{proof}
If $\mathcal B=0$, every component vanishes almost
everywhere, so $g=\mu$ with $J=0$ satisfies the claim.
Suppose that $\mathcal B>0$.

Since $N\geq1$ and $\rho_u\geq\rho_{\min}$,
Theorem 2 gives
\[
\|f-g\|_{L^2([0,1]^p)}
\leq
C N^{-\rho_{\min}}\mathcal B.
\]
Let $C_0=\max\{1,C\}$ and choose
\[
N
=
\left\lceil
(C_0/\varepsilon)^{1/\rho_{\min}}
\right\rceil.
\]
Then $C N^{-\rho_{\min}}\leq\varepsilon$, and hence
\[
\|f-g\|_{L^2([0,1]^p)}
\leq\varepsilon\mathcal B.
\]

Let $K=K_{\max}$. Since each $u\in\mathcal U$
satisfies $1\leq|u|\leq K$, the number of possible
index sets is bounded by
\[
|\mathcal U|
\leq
\sum_{d=1}^{\min(K,p)}\binom pd
\leq Kp^K.
\]
Also, $N+k+1\leq(k+2)N$ because $N\geq1$.
Therefore,
\[
\begin{aligned}
J
&\leq
\sum_{u\in\mathcal U}(N+k+1)^{d_u}\\
&\leq
|\mathcal U|(N+k+1)^K\\
&\leq
K(k+2)^Kp^K N^K.
\end{aligned}
\]
The choice of $N$ satisfies
\[
N
\leq
2C_0^{1/\rho_{\min}}
\varepsilon^{-1/\rho_{\min}},
\]
so
\[
J
=
O\!\left(
p^K\varepsilon^{-K/\rho_{\min}}
\right).
\]

For the uniform assertion, suppose that
$\overline C\geq1$ bounds all componentwise
approximation constants independently of $p$.
We may choose $C_0=\overline C$.
If $\rho_{\min}\geq\rho_*>0$, then
\[
N
\leq
2\overline C^{1/\rho_*}
\varepsilon^{-1/\rho_{\min}}.
\]
Consequently,
\[
J
\leq
K\{2(k+2)\}^K
\overline C^{K/\rho_*}
p^K\varepsilon^{-K/\rho_{\min}}.
\]
The implied constant therefore depends only on
$k$, $K$, $\overline C$, and $\rho_*$,
but not on $p$ or $\varepsilon$.
\end{proof}


\section{Additional Numerical Results and Implementation Details}
\label{sec:supp_sim}

\subsection{Representative Surface Fits}
\label{subsec:supp_surfaces}

\begin{sidewaysfigure}
    \centering
    \includegraphics[width=\textheight]{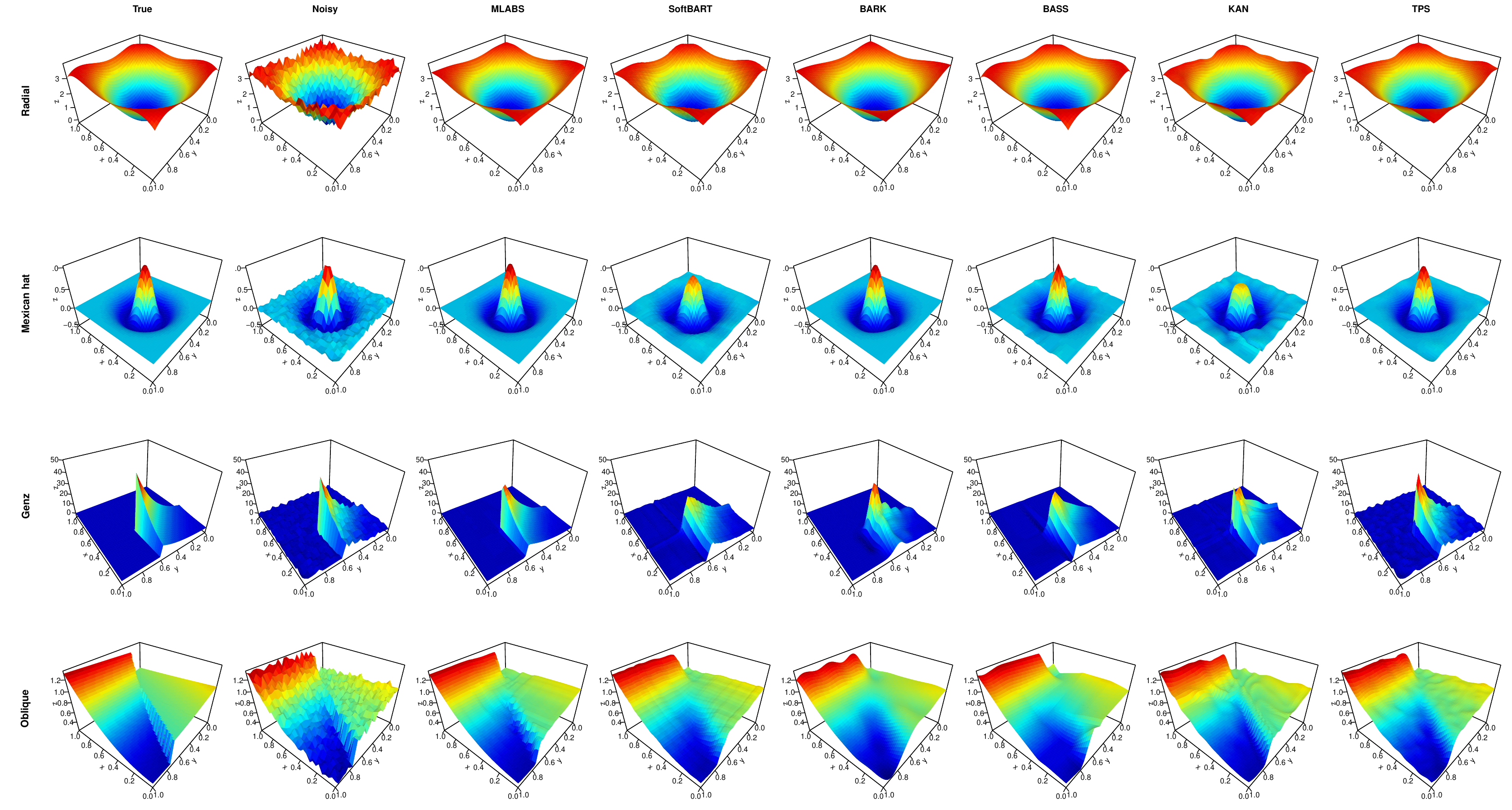}
    \caption{
    Representative surface fits for the four test functions at
    $\mathrm{RSNR}=5$. Rows correspond, from top to bottom, to the
    Radial, Mexican hat, Genz, and Oblique functions. Columns show the
    true surface, a noisy realization, and the fitted surfaces from
    MLABS, SoftBART, BARK, BASS, KAN, and TPS.
    }
    \label{fig:supp_surface_fits}
\end{sidewaysfigure}

Figure~\ref{fig:supp_surface_fits} shows representative fitted surfaces
for the four two-dimensional test functions at $\mathrm{RSNR}=5$.
The figure complements the predictive RMSE comparisons in the main
paper by illustrating how the fitted surfaces represent smooth,
localized, and discontinuous features.

\subsection{Hyperparameter Settings and Implementation Details}
\label{subsec:supp_tuning}

Hyperparameters were selected by five-fold cross-validation over the
candidate settings listed in Table~\ref{tab:hyper}. Unless otherwise
stated, tuning was performed separately for each simulation setting.

All competing methods were implemented using publicly available
software. MARS was fitted using the R package \pkg{earth}, Bayesian
MARS using \pkg{BASS} \citep{francom2020bass}, BARK using
\pkg{bark}, SVM using \pkg{e1071} \citep{meyer2015support},
neural networks using \pkg{keras3}, random forests using
\pkg{randomForest}, XGBoost using \pkg{xgboost}, BART using
\pkg{BayesTree}, SoftBART using \pkg{SoftBart}
\citep{linero2022softbart}, and FlexBART using \pkg{FlexBart}.
KAN was implemented in Python using the \pkg{pykan} package.

The neural network consisted of three fully connected hidden layers
with 64, 64, and 32 units and ReLU activation functions. For KAN, we
considered shallow, deep, and wide architectures. Let $p$ denote the
input dimension and define
\[
h_1=\min\{\max(2p,8),32\},
\qquad
h_2=\min\{\max(4p,16),64\}.
\]
The corresponding widths were $(p,h_1,1)$,
$(p,h_1,h_1,1)$, and $(p,h_2,1)$, respectively. Thus, for the
two-dimensional surface experiments these reduce to
$(2,8,1)$, $(2,8,8,1)$, and $(2,16,1)$.

For MLABS, the rate parameter of the gamma prior on the total mass
$M$ was fixed at $b_\gamma=1$, and the inverse-gamma prior on
$\sigma^2$ used $r=R=0.01$. The remaining tuning parameters control
the spline dictionary, interaction structure, model size, and
coefficient regularization. An additional multiplicative coefficient
scale used in the implementation is denoted by $m_b$ in
Table~\ref{tab:hyper}.

Each MLABS fit was based on $200{,}000$ MCMC iterations. The first
$100{,}000$ iterations were discarded as burn-in, and every tenth
post-burn-in draw was retained for posterior inference.

\clearpage
\begingroup
\small
\setlength{\tabcolsep}{4pt}
\renewcommand{\arraystretch}{1.12}
\setlength{\LTleft}{0pt}
\setlength{\LTright}{0pt}
\setlength{\LTcapwidth}{\textwidth}
\begin{longtable}{
|>{\raggedright\arraybackslash}p{0.14\textwidth}
|>{\raggedright\arraybackslash}p{0.30\textwidth}
|>{\raggedright\arraybackslash}p{\dimexpr0.56\textwidth-6\tabcolsep-4\arrayrulewidth\relax}|}
\caption{
Hyperparameters and candidate values considered in the numerical
experiments. For KAN,
$h_1=\min\{\max(2p,8),32\}$ and
$h_2=\min\{\max(4p,16),64\}$.
}
\label{tab:hyper}\\
\hline
\textbf{Method} & \textbf{Parameter} & \textbf{Values considered} \\
\hline
\endfirsthead
\multicolumn{3}{l}{\tablename~\thetable\ (continued)} \\
\hline
\textbf{Method} & \textbf{Parameter} & \textbf{Values considered} \\
\hline
\endhead
\hline
\multicolumn{3}{r}{Continued on next page} \\
\endfoot
\hline
\endlastfoot
\textbf{MLABS}
& Allowed spline-degree set, $S$
& $\{2\}$, $\{3\}$, $\{0,1\}$, $\{0,2\}$,
  $\{1,2\}$, $\{2,3\}$, $\{0,1,2\}$ \\
\hhline{|~|-|-|}

& Maximum interaction order, $K_{\max}$
& $2$, $3$ \\
\hhline{|~|-|-|}

& Boundary-expansion ratio, $E$
& $0.1$, $1$, $2$, $3$ \\
\hhline{|~|-|-|}

& Gamma-prior shape parameter for $M$, $a_\gamma$
& $5$, $10$, $30$, $50$ \\
\hhline{|~|-|-|}

& Coefficient-prior scale factor, $c_{\mathrm{shrink}}$
& $1$, $3$, $5$, $20$ \\
\hhline{|~|-|-|}

& Model-size penalty, $\lambda_J$
& $0$, $10$ \\
\hhline{|~|-|-|}


& Interaction-order prior weights, $\boldsymbol w_K$
& For $K_{\max}=2$: $(0.3,0.7)$, $(0.5,0.5)$, $(0,1)$;
  for $K_{\max}=3$: $(1/3,1/3,1/3)$ \\
\hline

\textbf{SBART}
& Number of trees
& $20$, $50$, $200$ \\
\hline

\textbf{FlexBART}
& Number of trees
& $20$, $50$, $200$ \\
\hline

\textbf{BART}
& Sigma prior, $(\nu,q)$
& $(3,0.90)$, $(3,0.99)$, $(10,0.75)$ \\
\hhline{|~|-|-|}

& Number of trees, $m$
& $50$, $200$ \\
\hhline{|~|-|-|}

& $\mu$-prior parameter, $k$
& $1$, $2$, $3$, $5$ \\
\hline

\textbf{BARK}
& Prior type for scale parameters
& ``e'', ``d'', ``se'', ``sd'' \\
\hline

\textbf{BASS}
& Spline degree
& $1$, $2$, $3$ \\
\hhline{|~|-|-|}

& Maximum interaction order
& $1$, $2$, $3$ \\
\hline

\textbf{MARS}
& Maximum number of terms in the pruned model
& $2$, $12$, $23$, $34$, $45$, $56$, $67$, $78$, $89$, $100$ \\
\hhline{|~|-|-|}

& Maximum interaction order
& $1$, $2$, $3$ \\
\hline

\pagebreak[4]
\textbf{RF}
& Number of variables considered at each split, $m_{\mathrm{try}}$
& 10 candidate values generated using \pkg{caret} with
  \texttt{tuneLength=10} \\
\hline

\textbf{SVM}
& Regularization constant, $C$
& $0.01$, $0.1$, $1$, $5$, $10$, $100$ \\
\hhline{|~|-|-|}

& Kernel parameter, $\gamma$
& $0.5$, $1$, $2$, $3$, $4$ \\
\hline

\textbf{NN}
& Hidden-layer architecture
& $(64,64,32)$ with ReLU activation \\
\hhline{|~|-|-|}

& Learning rate
& $0.001$, $0.005$, $0.01$, $0.05$, $0.1$ \\
\hline

\textbf{XGB}
& Maximum number of boosting iterations
& $250$, $500$, $1000$ \\
\hhline{|~|-|-|}

& Maximum tree depth
& $4$, $8$, $12$ \\
\hhline{|~|-|-|}

& Learning rate
& $0.05$, $0.1375$, $0.225$, $0.3125$, $0.40$ \\
\hhline{|~|-|-|}

& Minimum child weight
& $1$, $10$ \\
\hhline{|~|-|-|}

& Column subsampling ratio
& $0.7$, $1$ \\
\hhline{|~|-|-|}

& Minimum loss reduction, $\gamma$
& $0.1$ (fixed) \\
\hhline{|~|-|-|}

& Observation subsampling ratio
& $0.8$ (fixed) \\
\hline

\textbf{KAN}
& Shallow architecture
& $(p,h_1,1)$, grid $=5$, $k=3$,
  learning rate $=0.001$, steps $=300$ \\
\hhline{|~|-|-|}

& Deep architecture
& $(p,h_1,h_1,1)$, grid $=5$, $k=3$,
  learning rate $=0.001$, steps $=500$ \\
\hhline{|~|-|-|}

& Wide architecture
& $(p,h_2,1)$, grid $=10$, $k=3$,
  learning rate $=0.0005$, steps $=500$ \\
\hhline{|~|-|-|}

& Regularization parameter, $\lambda$
& $10^{-4}$ (fixed) \\

\end{longtable}
\endgroup

\subsection{Computation Time}
\label{subsec:supp_runtime}

We compare computation time in the high-dimensional Friedman~1
experiment. Table~\ref{tab:supp_runtime} reports the average elapsed
time, in seconds, across predictor dimensions and simulation
replications for each noise level. The methods included in the table
were evaluated in the same computing environment.

For KAN, at each combination of $p$ and $\sigma^2$, we used the
configuration corresponding to the reported predictive result and computed
its mean elapsed time across simulation replications. The reported
computation time for each noise level was then obtained by averaging these
mean runtimes over the predictor dimensions. KAN was run entirely on CPU
without GPU acceleration.

MLABS requires additional computation for trans-dimensional posterior
sampling and structural updates. Nevertheless, its average runtime is
of the same order as those of BART, SoftBART, and FlexBART in this
experiment, and is substantially smaller than those of BARK and
XGBoost. KAN requires more computation than MLABS and the Bayesian
tree-based methods considered here, but remains considerably faster
than BARK and XGBoost.

\begin{center}
\small
\captionof{table}{
Average computation time (in seconds) in the high-dimensional
Friedman~1 experiment, averaged across predictor dimensions and
simulation replications for each noise level.
}
\label{tab:supp_runtime}
\setlength{\tabcolsep}{3.5pt}

\begin{tabular}{lrrrrrr}
\toprule
Noise & BARK & BART & BASS & FlexBART & MARS & MLABS \\
\midrule
$\sigma^2=1$
& 1342.7 & 101.7 & 185.6 & 82.0 & 40.7 & 122.9 \\
$\sigma^2=10$
& 1304.4 & 76.2 & 181.9 & 86.2 & 44.3 & 132.2 \\
\midrule
Noise & NN & RF & SVM & SoftBART & XGBoost & KAN \\
\midrule
$\sigma^2=1$
& 6.4 & 14.7 & 1.1 & 110.5 & 1171.7 & 320.5 \\
$\sigma^2=10$
& 6.4 & 15.2 & 1.1 & 111.7 & 1239.4 &  526.1 \\
\bottomrule
\end{tabular}

\end{center}

\subsection{Model Complexity of MLABS}
\label{subsec:supp_complexity}

\begin{figure}[ht!]
\centering
\includegraphics[width=0.65\textwidth]{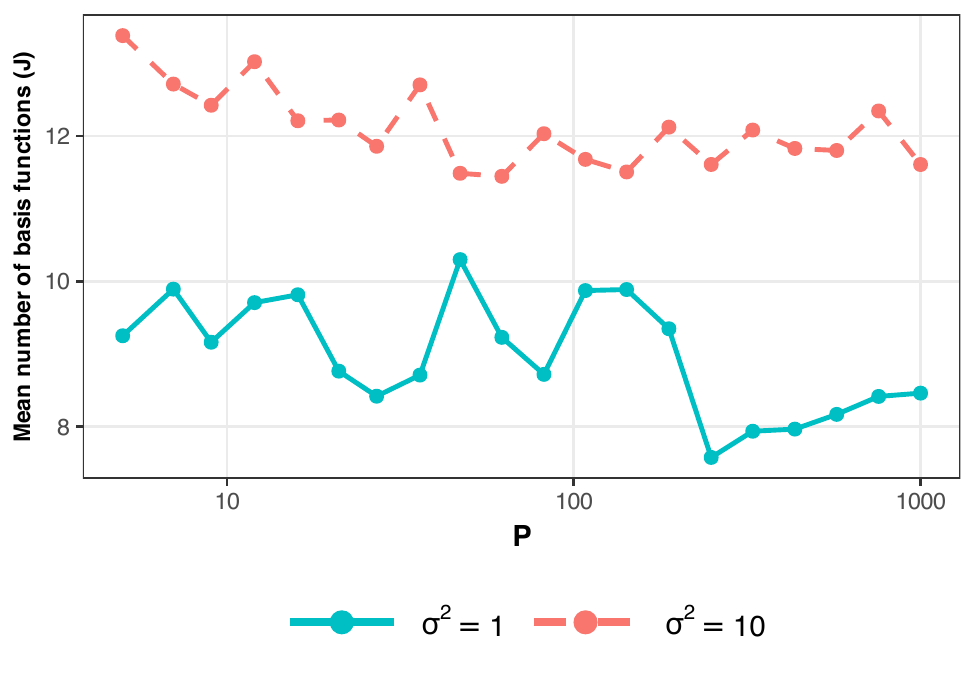}
\caption{
Posterior mean number of basis functions in MLABS as a function of the
number of predictors $p$ in the high-dimensional Friedman~1 experiment.
Results are shown separately for $\sigma^2=1$ and $\sigma^2=10$.
}
\label{fig:supp_mean_basis}
\end{figure}

We also examine how the size of the fitted MLABS expansion changes
with the ambient dimension. Figure~\ref{fig:supp_mean_basis} shows the
posterior mean number of basis functions as a function of $p$ in the
high-dimensional Friedman~1 experiment.

Across the dimensions considered, the posterior mean number of basis
functions remains approximately between 7 and 13 under both noise
levels, despite the increase from $p=5$ to $p=1000$. This result
indicates that the fitted expansion remains relatively sparse as the
number of irrelevant predictors increases.

\end{appendices}

\end{document}